\documentclass[aps,twocolumn,pre,floatfix]{revtex4-2}

\usepackage{xcolor,hyperref}
\hypersetup{
   colorlinks,
   linkcolor={blue!50!black},
   citecolor={blue!50!black},
   urlcolor={blue!80!black}
}

\usepackage{epsfig,amsmath}
\usepackage{bm}
\usepackage{soul,ulem}
\usepackage{hyperref}
\usepackage{footmisc}
\usepackage{wrapfig}
\DeclareMathAlphabet{\mathitbf}{OML}{cmm}{b}{it}

\newcommand{\zerovector}{\bm{0}}

\renewcommand{\=}{\!=\!}

\newcommand{\ket}[1]{|#1\rangle}

\newcommand{\Svv}{\mathcal S}

\newcommand{\uv}{\mathitbf u}
\newcommand{\xv}{\mathitbf x}
\newcommand{\Xv}{\mathitbf X}

\newcommand{\nv}{\mathitbf n}

\newcommand{\sFrac}[2]{{\textstyle\frac{#1}{#2}}}
\newcommand{\dbar}{{\,\mathchar'26\mkern-12mu d}}

\DeclareMathAlphabet\mathbfcal{OMS}{cmsy}{b}{n}
\begin{document}

\title{The strain-stiffening critical exponents in polymer networks and their universality}
\author{Zibin Zhang}
\author{Eran Bouchbinder$^2$}
\author{Edan Lerner$^1$}
\affiliation{$^1$Institute for Theoretical Physics, University of Amsterdam, Science Park 904, 1098 XH Amsterdam, the Netherlands\\
$^2$Chemical and Biological Physics Department, Weizmann Institute of Science, Rehovot 7610001, Israel}

\begin{abstract}
Disordered athermal biopolymer materials, such as collagen networks that constitute a major component in extracellular matrices
and various connective tissues, are initially soft and compliant but stiffen dramatically under strain. Such network materials are topologically sub-isostatic and feature strong rigidity scale separation between the bending and stretching response of the constituent polymer fibers. Recently, a comprehensive scaling theory of the athermal strain-stiffening phase transition has been developed, providing predictions for all critical exponents characterising the transition in terms of the distance to the critical strain and of the small rigidity scales ratio. Here, we employ large-scale computer simulations, at and away from criticality, to test the analytic predictions. We find that all numerical critical exponents are in quantitative agreement with the analytically-predicted ones. Moreover, we find that all predicted exponents remain valid whether the driving strain is shear, i.e., volume-preserving, or dilation, and independent of the degree of the network's sub-isostaticity, thus establishing the universality of the strain-stiffening phase transition with respect to the symmetry of the driving strain and the network's topology.
\end{abstract}

\maketitle

\section{I\lowercase{ntroduction}}

\begin{figure}[ht!]
  \includegraphics[width = 0.5\textwidth]{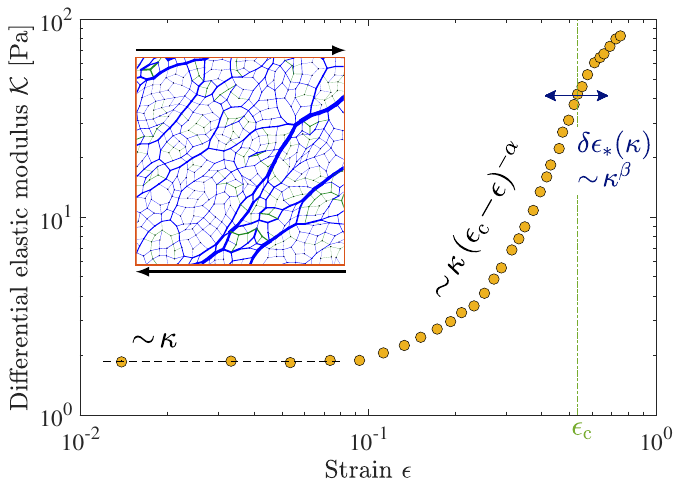}
  \caption{\small {\bf The strain-stiffening transition in athermal polymer networks.} Experimental data (circles) obtained for reconstituted collagen networks under shear (extracted from the dataset in Fig.~2E of~\cite{licup2015stress} that corresponds to a protein concentration of $0.90$ mg/mL and a polymerization temperature of $37^\circ$ C). Shown is the differential elastic modulus, denoted by ${\cal K}$, as a function of the applied strain, denoted by $\epsilon$, on a double-logarithmic scale. In the absence of strain, or when $\epsilon$ is very small, the network is soft and its modulus is proportional to $\kappa$ (horizontal dashed line), a dimensionless measure of the small bending-to-stretching ratio of individual polymer fibers. As $\epsilon$ is increased, the network undergoes dramatic power-law stiffening in terms of the strain distance from a critical strain $\epsilon_{\rm c}$ ($\alpha\!=\!3/2$ is theoretically predicted, see Eq.~\eqref{eq:away_from_critical_a}). The critical strain $\epsilon_{\rm c}$ (vertical dashed-dotted line) is accompanied by a characteristic strain scale $\delta\epsilon_*(\kappa)$ (double arrow), which follows a scaling relation with $\kappa$ ($\beta\!=\!2/3$ is theoretically predicted, see Eq.~\eqref{eq:typical_strain}). (inset) A zoom in on a two-dimensional computer network, composed of nodes and bonds (see details in the text), driven to the critical strain in the absence of bending energy ($\kappa\!=\!0$). Here, as in the experimental data in the main panel, shear straining is applied, indicated by the two oppositely oriented black arrows. Bonds in blue/green experience stretching/compression, where the bond's thickness represents the magnitude. Shearing-induced anisotropy is evident.}
  \label{fig:fig1}
\end{figure}

Natural and manmade network materials, composed of interacting fibers/filaments, are widespread~\cite{picu2022network}. These materials achieve quite remarkable mechanical properties and functionalities with relatively low volume fractions, i.e., without approaching the space-filling limit. The latter is particularly important in living systems, where naturally emerging biomaterials avoid the high costs associated with protein-based self-assembly. In many cases, network materials are initially soft and compliant, but stiffen dramatically in response to mechanical strain~\cite{kees_nature_2005,erk2010strain,robbie_nature_physics_2016,vatankhah2018chameleon}. That is, their elastic moduli increase by several orders of magnitude over rather narrow intervals of strain, a generic phenomenon termed strain-stiffening, see Fig.~\ref{fig:fig1}. Consider, for example, tissues such as skin, which are soft in their undeformed state, but stiffen significantly under strain to avoid damage~\cite{picu2022network,kees_nature_2005}. In a biological context, strain-stiffening is not only important for self-protection, but also for cell-cell communication, cell fate and morphology, and tissue development~\cite{Janmey_soft_matter_2007,winer2009non,das2016stress,han2018cell,wang2023strain}.

The vast majority of network materials that feature a strain-stiffening transition are intrinsically disordered. Among these, one can distinguish between disordered networks composed of semiflexible polymers, where significant thermal bending fluctuations --- hence entropic effects --- play important roles, and disordered athermal polymer networks, where thermal fluctuations are negligible. In the context of synthetic/manmade materials, disordered athermal networks include cellulose and polymeric fiber networks in paper, nonwoven fabrics and various textiles~\cite{picu2022network}. A prominent example in a biological context is that of collagen networks, which are predominantly athermal~\cite{picu2022network,robbie_nature_physics_2016}. Disordered collagen networks are present everywhere in our bodies~\cite{picu2022network,robbie_nature_physics_2016}, being a major component in extracellular matrices and in various connective tissues (e.g., cartilage, tendons and ligaments), and hence are of prime importance.

In addition to their disordered nature, athermal polymer networks that exhibit a strain-stiffening transition are characterized by two other generic features; first, these disordered networks are topologically sub-isostatic, i.e., their degree of connectivity (average network coordination) is below the Maxwell threshold~\cite{maxwell_1864}. This implies that such networks can be deformed with no energetic cost, i.e., without involving any stretching or compression of the constituent polymers to leading order, due to the existence of zero (floppy) modes~\cite{phonon_gap_2012}. Consequently, from a topological perspective, these networks are floppy, featuring vanishing elastic moduli. However, realistic athermal polymer networks feature small, yet finite, elastic moduli in their unstrained state, which imply that they are stabilized by additional weak interactions on top of polymer fibers stretching/compression.

Indeed, the second generic property of such systems is that the polymer fibers feature finite bending rigidity, preventing them from being floppy, but rather posses a small elastic resistance in the absence of strain. As the resistance of the polymer fibers to elongation/shortening (stretching/compression) is much larger than to bending, these networks generically exhibit strong rigidity/stiffness scale separation. Understanding how these generic features endow athermal polymer networks with intriguing universal properties is an important challenge.

Strain-stiffening in athermal polymer networks has been studied through various experimental, theoretical and computational approaches~\cite{kees_nature_2005,Janmey_soft_matter_2007,erk2010strain,RevModPhys.86.995,gustavo_pre_2014,robbie_pre_2018,merkel_pnas_2019,mackintosh_pre_2016,Rens_JPCB_2016,robbie_nature_physics_2016,sharma2016strain,wouter_pre_2017,robbie_pre_2018,rens2019theory,maha_prl_2008,mackintosh_prl_2019,arzash2021shear,fred_arXiv_2022,chen2023effective,freddy_mac_prl_2024}. Progress has been made in a series of important works that identified and highlighted the critical nature of the strain-stiffening transition~\cite{licup2015stress,robbie_nature_physics_2016}. That is, these works indicated that strain-stiffening in athermal polymer networks is a strain-driven phase transition. Very recently, building on this physical picture, a comprehensive scaling theory of the transition has been developed~\cite{lerner2023scaling}, offering analytic predictions for all the scaling exponents characterizing it. In this work, we numerically test the recently predicted strain-stiffening scaling relations, the values of the critical exponents and their degree of universality. We find great quantitative agreement with all analytic predictions, and also demonstrate their universality with respect to the symmetry of the driving strain and the degree of iso-staticity.

\section{T\lowercase{heoretical predictions}}

We considered disordered elastic networks composed of nodes connected by bonds, representing the polymer fibers, whose topology
is quantified by the average connectivity (bonds per node) $z\!<\!z_{\rm c}\=2\dbar$, where the latter is the Maxwell
rigidity criterion in $\dbar$ spatial dimensions~\cite{maxwell_1864}. That is, these networks are sub-isostatic, characterized by $\delta{z}\!\equiv\!z_{\rm c}-z\!>\!0$. The bonds feature stretching/compression rigidity that is much larger than their bending rigidity, giving rise to a dimensionless rigidity scales (bending-to-stretching) ratio $\kappa\!\ll\!1$ (the latter involves a lengthscale, the average bond length, see {\color{blue}{\it Appendix}}). The networks are driven by an applied strain tensor that is parameterized by an amplitude $\epsilon$ and whose symmetry (e.g., shear vs.~dilation) remains unspecified for now. In the absence of bending interactions, $\kappa\=0$, sub-isotatic networks undergo a sharp rigidity transition at a critical strain $\epsilon_{\rm c}$, upon which the elastic modulus ${\cal K}$ associated with $\epsilon$ jumps discontinuously from zero to a finite value~\cite{wouter_pre_2017,rens2019theory}. In the presence of weak bending interactions, $\kappa\!\ll\!1$, such networks undergo a strong --- yet continuous --- strain-stiffening transition as a function of $\epsilon$, see Fig.~\ref{fig:fig1}. The challenge is to understand this driven phase transition in terms of $\delta{z}$, $\kappa$ and $\Delta\epsilon\!\equiv\!\epsilon_{\rm c}-\epsilon$, and to elucidate its dependence on the imposed straining symmetry.

The strain-stiffening transition is manifested through various physical observables. At the macroscopic scale, the network is characterized by its energy $U$ and its derivatives with respect to the strain $\epsilon$, most notably the stress $\sigma$ (first derivative) and the differential modulus ${\cal K}$ (second derivative), see {\color{blue}{\it Appendix}} for explicit definitions. The behavior of ${\cal K}(\epsilon,\kappa)$ (at a given $\delta{z}$) is illustrated in Fig.~\ref{fig:fig1} using experimental data for a reconstituted collagen network~\cite{licup2015stress}. It is observed that an initially soft network dominated by bending interactions, ${\cal K}(\epsilon,\kappa)\!\sim\!\kappa$ for $\epsilon\!\ll\!\epsilon_{\rm c}$, undergoes power-law, apparently singular stiffening with increasing strain $\epsilon$ for $\epsilon_{\rm c}-\epsilon\!\gg\!\delta\epsilon_*(\kappa)$, where $\delta\epsilon_*(\kappa)\!\ll\!\epsilon_{\rm c}$ is a characteristic strain scale that quantifies the proximity to the critical strain $\epsilon_{\rm c}$. For $\epsilon_{\rm c}-\epsilon\!\ll\!\delta\epsilon_*(\kappa)$, i.e., near the critical point, the apparently singular stiffening is regularized. In addition to these macroscopic observables, the network is also characterized by its internal deformation and state of disorder, which give rise to the macroscopic behaviors.

Very recently, a comprehensive scaling theory of the strain-stiffening transition has been developed, offering analytic predictions for all of the above-mentioned physical observables~\cite{lerner2023scaling}. Here, we briefly discuss the conceptual framework underlying the theory and its main predictions. As explained above, a central quantity is the network's energy $U(\epsilon,\kappa)$. The theory first focuses on the critical state, $\epsilon\=\epsilon_{\rm c}$, and obtains $U(\epsilon_{\rm c},\kappa)$ as follows; the entire network is decomposed into a stiff sub-network, characterized by the stiff stretching/compression interactions, and a soft sub-network, characterized by the soft/weak bending interactions, that are coupled at the network's nodes. The basic idea is to use the smallness of $\kappa\!\ll\!1$ to derive the scaling properties of $U(\epsilon_{\rm c},\kappa)$ by constructing it in two steps. First, the stiff sub-network is strained to the critical strain $\epsilon\=\epsilon_{\rm c}$, where a finite elastic modulus emerges, yet featuring $U_{\mbox{\tiny stiff}}(\epsilon_{\rm c},\kappa\=0)\=0$. Second, the soft sub-network is added at $\epsilon\=\epsilon_{\rm c}$, giving rise to nodal displacements of characteristic size $u_*$, accompanied by $U_{\mbox{\tiny stiff}}(\epsilon_{\rm c},\kappa)\!>\!0$ and $U_{\mbox{\tiny soft}}(\epsilon_{\rm c},\kappa)\!>\!0$.

The introduction of the soft sub-network gives rise to net/resultant nodal forces of scale $F_{\mbox{\tiny soft}}\!\sim\!\kappa$, which are balanced by the nodal forces generated by the stiff sub-network. The crucial point is that the stiff sub-network's response to the perturbation introduced by the soft sub-network is dominated by a fourth-order anharmonicity, $U_{\mbox{\tiny stiff}}(\epsilon_{\rm c},\kappa)\!\sim\!u^4$, where $u$ is a characteristic nodal displacement. The origin of this striking result is that while some of the zero (floppy) modes of the stiff sub-network are destroyed at the critical strain $\epsilon_{\rm c}$, an extensive number of them persist, i.e., they are uncoupled to the strain, and dominate the response to the isotropic perturbation introduced by the soft sub-network. The energy of these zero modes vanishes to order $u^2$. The cubic contribution $\sim u^3$ vanishes to ensure stability and hence the quartic contribution dominates. Mechanical equilibrium requires $F_{\mbox{\tiny stiff}}\!\sim\!u^3\=F_{\mbox{\tiny soft}}\!\sim\!k$, implying that the nodal displacements induced by the introduction of the soft sub-network satisfy
\begin{equation}
\label{eq:displacement}
u_*(\kappa) \sim \kappa^{1/3} \ .
\end{equation}

The fundamental result in Eq.~\eqref{eq:displacement} has major implications. First, it implies that $U_{\mbox{\tiny stiff}}(\epsilon_{\rm c},\kappa)\!\sim\!u_*^4\!\sim\!\kappa^{4/3}$. Since the soft sub-network undergoes a $\kappa$-independent deformation for sufficiently small $\kappa$, it satisfies $U_{\mbox{\tiny soft}}(\epsilon_{\rm c},\kappa)\!\sim\!\kappa$, which implies
\begin{equation}
\label{eq:energy}
   U(\epsilon_{\rm c},\kappa) = U_{\mbox{\tiny stiff}}(\epsilon_{\rm c},\kappa) + U_{\mbox{\tiny soft}}(\epsilon_{\rm c},\kappa) \sim U_{\mbox{\tiny soft}}(\epsilon_{\rm c},\kappa) \sim \kappa \ .
\end{equation}
For the stress, one obtains~\cite{lerner2023scaling}
\begin{equation}
\label{eq:stress}
\sigma(\epsilon_{\rm c},\kappa) \sim \kappa^{2/3} \ ,
\end{equation}
which is dominated by bond stretching/compression, $f_{\mbox{\tiny stiff}}\!\gg\!f_{\mbox{\tiny soft}}\!\sim\!\kappa$ (where bond force $f$ should be distinguished from net/resultant nodal forces, which satisfy $F_{\mbox{\tiny stiff}}\=F_{\mbox{\tiny soft}}$), in striking contrast to the energy in Eq.~\eqref{eq:energy}, which is dominated by bond bending. Yet another implication of Eq.~\eqref{eq:displacement} is $d{\cal K}(\epsilon_{\rm c},\kappa)/d\epsilon\!\sim\!\kappa^{-2/3}$~\cite{lerner2023scaling}, which shows how $\kappa\!>\!0$ regularizes the singular behavior of the differential modulus ${\cal K}$ as the critical strain $\epsilon_{\rm c}$ is approached.

The behavior of $d{\cal K}(\epsilon_{\rm c},\kappa)/d\epsilon$ is intimately related to the disordered nature of the elastic networks under discussion. Another fundamental manifestation of the disorder of these systems is that their deformation is highly non-affine. That is, the deformation of such disordered systems does not follow the globally applied strain, but rather reveals additional `relaxational deformation' termed non-affine displacements of characteristic magnitude $u_{\mbox{\tiny n.a.}}$, which satisfies~\cite{lerner2023scaling}
\begin{equation}
\label{eq:na_displacements_squared}
u_{\mbox{\tiny n.a.}}^2(\epsilon_{\rm c},\kappa) \sim \kappa^{-2/3}\ .
\end{equation}
Equations~(\ref{eq:displacement})-(\ref{eq:na_displacements_squared}) offer predictions for the $\kappa$ dependence of basic quantities at the critical strain $\epsilon_{\rm c}$. The prediction $d{\cal K}(\epsilon_{\rm c},\kappa)/d\epsilon\!\sim\!\kappa^{-2/3}$ immediately implies that ${\cal K}(\epsilon,\kappa)-{\cal K}(\epsilon_{\rm c})\!\sim\!\kappa^{-2/3}(\epsilon-\epsilon_{\rm c})$ for $\epsilon$ sufficiently close to $\epsilon_{\rm c}$.
\begin{figure*}[ht!]
 \includegraphics[width = 1\textwidth]{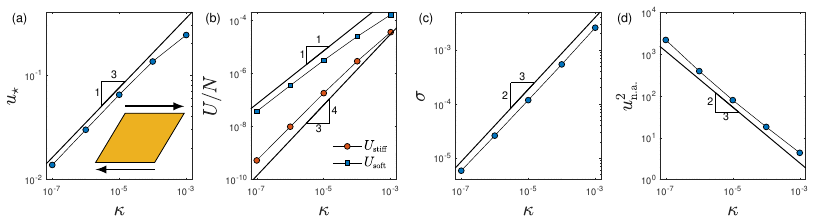}
  \caption{\small {\bf Numerical validation of the strain-stiffening critical exponents under shear.} The physical quantities theoretically predicted in Eqs.~(\ref{eq:displacement})-(\ref{eq:na_displacements_squared}) are calculated in 2D computer simulations of disordered elastic networks with $\delta{z}\!=\!0.5$ (see text and {\color{blue}{\it Appendix}} for details) under shear strain $\gamma$ (see inset in panel(a)) and plotted as a function of $\kappa$ at the critical strain-stiffening state $\gamma_{\rm c}$. (a) The characteristic nodal displacement $u_*(\gamma_{\rm c},\kappa)$. (b) The energies $U_{\mbox{\tiny stiff}}(\gamma_{\rm c},\kappa)$ and $U_{\mbox{\tiny soft}}(\gamma_{\rm c},\kappa)$ of the stiff and soft sub-networks, respectively. (c) The shear stress $\sigma(\gamma_{\rm c},\kappa)$. (d) The non-affine displacements squared $u_{\mbox{\tiny n.a.}}^2(\gamma_{\rm c},\kappa)$. The values of the critical exponents, indicated by the power-law triangles, are all in excellent quantitative agreement with the analytic predictions in Eqs.~(\ref{eq:displacement})-(\ref{eq:na_displacements_squared}).}
  \label{fig:fig2}
\end{figure*}

To obtain predictions away from the critical strain, it is essential to quantify what `sufficiently close to $\epsilon_{\rm c}$' actually means. The theory of~\cite{lerner2023scaling} suggests that there exists a characteristic strain scale $\delta\epsilon_*(\kappa)$ around $\epsilon_{\rm c}$, which allows to quantify the proximity to $\epsilon_{\rm c}$ and satisfies
\begin{equation}
\label{eq:typical_strain}
    \delta\epsilon_*(\kappa) \sim \kappa^{2/3} \ ,
\end{equation}
see Fig.~\ref{fig:fig1}. The existence of a characteristic $\kappa$-dependent strain scale $\delta\epsilon_*(\kappa)$ near $\epsilon_{\rm c}$ allows to construct a scaling theory also away from the critical point in terms of the scaled variable $(\epsilon_{\rm c}-\epsilon)/\delta\epsilon_*(\kappa)$. This analysis gave rise to the following predictions~\cite{lerner2023scaling}
\begin{eqnarray}
\label{eq:away_from_critical_a}
    {\cal K}(\epsilon,\kappa)\! &\sim& \!\kappa\,(\epsilon_{\rm c}-\epsilon)^{-3/2} \quad\! \hbox{for} \quad \epsilon_{\rm c}-\epsilon\gg\delta\epsilon_*(\kappa)\ ,\\
\label{eq:away_from_critical_b}
    u_{\mbox{\tiny n.a.}}^2\!(\epsilon,\kappa) &\sim& (\epsilon_{\rm c}-\epsilon)^{-1} \qquad\, \hbox{for} \quad \epsilon_{\rm c}-\epsilon\gg\delta\epsilon_*(\kappa)\ .
\end{eqnarray}

The analytic predictions in Eqs.~(\ref{eq:displacement})-(\ref{eq:away_from_critical_b}) provide a comprehensive scaling theory of the strain-stiffening transition, including all of the major critical exponents. The theory also reveals the singular perturbation nature of the critical state, as manifested in various divergencies as $\kappa\!\to\!0$ (e.g., the susceptibility $du_*(\kappa)/d\kappa$ diverges as $\kappa^{-2/3}$ in this limit, see also {\color{blue}{\it Appendix}}). The derivation leading to the above predictions makes no reference to space dimensionality, to the symmetry of the strain tensor and to $\delta{z}$, and hence predicts that the latter appear only affect the pre-factors. Consequently, the theory predicts the scaling relations and critical exponents listed above are universal with respect to space dimensionality $\dbar$, the symmetry of the strain tensor and the network's topology quantified by $\delta{z}$.

\section{N\lowercase{umerical validation of the critical exponents under shear straining}}

The analytic predictions in Eqs.~(\ref{eq:displacement})-(\ref{eq:away_from_critical_b}) call for numerical validation. Some preliminary support to the predictions in Eqs.~(\ref{eq:away_from_critical_a})-(\ref{eq:away_from_critical_b}) is obtained through comparison to numerical results available in the literature (see~\cite{lerner2023scaling}, in particular Table 1 therein). This agreement not only provides support to the values of the scaling exponents in Eqs.~(\ref{eq:away_from_critical_a})-(\ref{eq:away_from_critical_b}), but also supports the validity the two-step procedure and the decomposition of the system into interacting stiff and soft sub-networks, where the latter is viewed as a (singular) perturbation on top of the former. This is the case because the results available in the literature were obtained by numerically straining $\kappa\!>\!0$ disordered networks, without invoking the two-step procedure and the aforementioned sub-networks decomposition. Yet, a direct and comprehensive test of the predictions in Eqs.~(\ref{eq:displacement})-(\ref{eq:na_displacements_squared}) at criticality, and of Eqs.~(\ref{eq:typical_strain})-(\ref{eq:away_from_critical_b}), away from criticality, are currently missing.
\begin{figure}[ht!]
 \includegraphics[width = 0.5\textwidth]{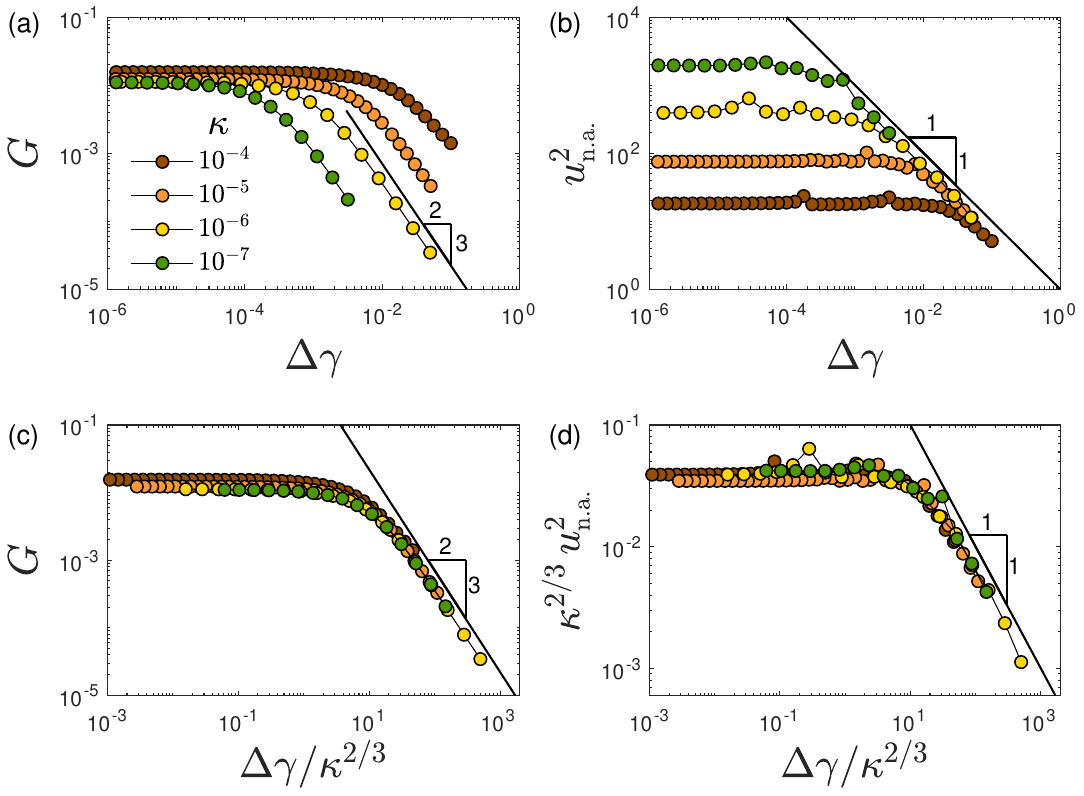}
  \caption{\small {\bf Numerical validation of the characteristic strain scale near the critical strain and of the strain-stiffening power-law scaling relations away from criticality under shear.} Numerical tests of the predictions in Eqs.~(\ref{eq:typical_strain})-(\ref{eq:away_from_critical_b}) under shear with $\delta{z}\!=\!0.5$. (a) The differential shear modulus $G(\Delta\gamma,\kappa)$ vs.~$\Delta\gamma\!=\!\gamma_{\rm c}-\gamma$ for various $\kappa$ values (see legend). (b) The non-affine displacements squared $u_{\mbox{\tiny n.a.}}^2(\Delta\gamma,\kappa)$ for various $\kappa$ values (see legend). (c) The same as panel (a), but with a rescaled $\Delta\gamma/\kappa^{2/3}$ x-axis, according to the prediction $\delta\epsilon_*(\kappa)\!\sim\!\kappa^{2/3}$ in Eq.~\eqref{eq:typical_strain} for the characteristic strain scale near the critical strain. Excellent data collapse for $\Delta\gamma/\kappa^{2/3}\!\gg\!1$ is observed (note the power-law triangle), as predicted by Eq.~\eqref{eq:away_from_critical_a}. (d) The same as panel (b), but for rescaled x- and y-axes, as indicated. Excellent data collapse is observed (note the power-law triangle), as predicted by Eqs.~(\ref{eq:na_displacements_squared}), (\ref{eq:typical_strain}) and~(\ref{eq:away_from_critical_b}).}
  \label{fig:fig3}
\end{figure}
\begin{figure*}[ht!]
 \includegraphics[width = 1\textwidth]{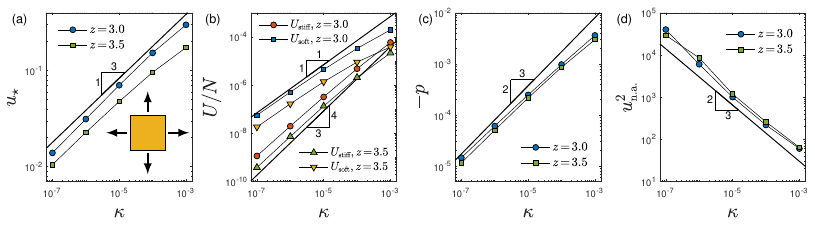}
  \caption{\small {\bf Numerical validation of the strain-stiffening critical exponents under dilation and universality with respect to $\delta{z}$.} The same as Fig.~\ref{fig:fig2}, but for dilation quantified by the volumetric strain amplitude $\eta$ (see inset in panel (a)) and two values of $\delta{z}$ (see legend). Note that all quantities are computed at the critical strain-stiffening strain $\eta_{\rm c}$ and that in panel (c) the hydrostatic tension $-p(\eta_{\rm c},\kappa)$ is plotted. The values of the critical exponents, indicated by the power-law triangles, are all in excellent quantitative agreement with the analytic predictions in Eqs.~(\ref{eq:displacement})-(\ref{eq:na_displacements_squared}).}
  \label{fig:fig4}
\end{figure*}

Numerically testing the analytic prediction is highly challenging, especially at criticality. The reason is that at $\epsilon_{\rm c}$ the network's energy is vanishingly small and its landscape is flat, dominated by quartic anharmonicity, as discussed above. Moreover, the strong rigidity scale separation implied by the smallness of $\kappa$ poses another challenge. We overcome these difficulties by performing large-scale computer simulations that realize the same two-step procedure invoked in the theoretical derivation~\cite{lerner2023scaling}.

In Fig.~\ref{fig:fig2}, we numerically test the predictions in Eqs.~(\ref{eq:displacement})-(\ref{eq:na_displacements_squared}) for the critical strain-stiffening transition in 2D ($\dbar\=2$) under shear straining. That is, the applied strain tensor corresponds to simple shear deformation (see inset in Fig.~\ref{fig:fig2}a and {\color{blue}{\it Appendix}}), which is parameterized by a strain amplitude $\gamma$. The presented results were obtained for disordered networks of $N\=6400$ nodes with $\delta{z}\=0.5$, averaged over an ensemble of a few tens of independent realizations of the disorder. Results at the critical strain $\gamma_{\rm c}$ are presented for the characteristic nodal displacement $u_*(\kappa)$ in panel (a), for the energies $U_{\mbox{\tiny stiff}}(\epsilon_{\rm c},\kappa)$ and $U_{\mbox{\tiny soft}}(\epsilon_{\rm c},\kappa)$ of the stiff and soft sub-networks, respectively, in panel (b), for the shear stress $\sigma(\epsilon_{\rm c},\kappa)$ in panel (c) and for the non-affine displacements squared $u_{\mbox{\tiny n.a.}}^2(\epsilon_{\rm c},\kappa)$ in panel (d). The $\kappa$ dependence of all of these quantities are in great quantitative agreement with the analytic predictions in Eqs.~(\ref{eq:displacement})-(\ref{eq:na_displacements_squared}), for sufficiently small $\kappa$.

We then set out to test the prediction for the characteristic strain scale in Eq.~\eqref{eq:typical_strain}, applied to shear straining, i.e., for $\delta\gamma_*(\kappa)$. The most natural way to achieve this is through testing Eqs.~(\ref{eq:typical_strain})-(\ref{eq:away_from_critical_b}), predicted to be valid away from criticality, i.e., for $\gamma_{\rm c}-\gamma\gg\delta\gamma_*(\kappa)$. The reason for this is that the predictions in Eqs.~(\ref{eq:away_from_critical_a})-(\ref{eq:away_from_critical_b}) involve the rescaled variable $\Delta\gamma/\delta\gamma_*(\kappa)\!\equiv\!(\gamma_{\rm c}-\gamma)/\delta\gamma_*(\kappa)$. In Fig.~\ref{fig:fig3}a, we present results for the differential shear modulus $G$ --- which is the relevant response quantity coupled to $\gamma$ (i.e., the relevant ${\cal K}$ in Eq.~\eqref{eq:away_from_critical_a} is $G$) --- as a function of $\Delta\gamma$ for various $\kappa$ values (see legend). These results were obtained by bringing the networks to the critical state at $\gamma_{\rm c}$ following the two-step procedure and then reducing the strain $\gamma$ (back-straining) to achieve the desired $\Delta\gamma$ values. Equations~(\ref{eq:typical_strain})-(\ref{eq:away_from_critical_a}) predict together that the different $G$ curves of Fig.~\ref{fig:fig3}a would collapse on a single power-law curve with a $-3/2$ exponent for $\Delta\gamma/\kappa^{2/3}\!\gg\!1$, once plotted against the rescaled variable $\Delta\gamma/\kappa^{2/3}$. This is done in Fig.~\ref{fig:fig3}c, revealing excellent agreement with the prediction, thus proving strong support to Eqs.~(\ref{eq:typical_strain})-(\ref{eq:away_from_critical_a}).

A similar procedure can be followed to test the predictions in Eqs.~\eqref{eq:typical_strain} and~\eqref{eq:away_from_critical_b}. To that aim, we plot $u_{\mbox{\tiny n.a.}}^2$ as a function of $\Delta\gamma$ for various $\kappa$ values in Fig.~\ref{fig:fig3}b. Equations~(\ref{eq:na_displacements_squared}), ({\ref{eq:typical_strain}) and~(\ref{eq:away_from_critical_b}) then predict that replotting these curves as $\kappa^{2/3}\,u_{\mbox{\tiny n.a.}}^2$ against the rescaled variable $\Delta\gamma/\kappa^{2/3}$ would result in a collapse onto a single curve, featuring a constant for $\Delta\gamma/\kappa^{2/3}\!\ll\!1$ and a power-law with a $-1$ exponent for $\Delta\gamma/\kappa^{2/3}\!\gg\!1$. Such a replotting, presented in Fig.~\ref{fig:fig3}d, reveals excellent agreement with the predictions, and with other simulational results~\cite{rens2019theory}. Note that a series of works~\cite{sharma2016strain,mackintosh_prl_2019,arzash2021shear,freddy_mac_expansion_pre_2022,fred_arXiv_2022,chen2023effective} persistently predicted $u_{\mbox{\tiny n.a.}}^2\!\sim\!(\Delta\gamma)^{-3/2}$, which is inconsistent with our findings. Taken together, Figs.~\ref{fig:fig2}-\ref{fig:fig3} provide strong support to all of the analytic strain-stiffening predictions in Eqs.~(\ref{eq:displacement})-(\ref{eq:away_from_critical_b}) under shear straining.

\section{U\lowercase{niversality with respect to the network's topology and the symmetry of the driving strain}}

As mentioned above, the predictions in Eqs.~(\ref{eq:displacement})-(\ref{eq:away_from_critical_b}) are expected to be independent of the symmetry of the strain tensor and the network's topology quantified by $\delta{z}$. In the previous section, the predictions have been verified under shear straining for a single value of $\delta{z}$.  Our goal here is to consider a different symmetry of the applied strain, specifically dilatational straining, and to vary $\delta{z}$ as well. Dilatational strains are particularly interesting, not just because they are very different from volume-preserving shear strains, but also because recent work~\cite{freddy_mac_expansion_pre_2022} suggested that strain-stiffening critical exponents might differ between shear and dilation.

In Fig.~\ref{fig:fig4}, we numerically test the predictions in Eqs.~(\ref{eq:displacement})-(\ref{eq:na_displacements_squared}) for the critical strain-stiffening transition in 2D ($\dbar\=2$) under dilatational (volumetric) straining. That is, the applied strain tensor corresponds to pure dilatational deformation (see inset in Fig.~\ref{fig:fig4}a and {\color{blue}{\it Appendix}}), which is parameterized by a strain amplitude $\eta$. The presented results were obtained for disordered networks of $N\=6400$ nodes with two $\delta{z}$ values (see legend), averaged over an ensemble of a few tens of independent realizations of the disorder. The results presented in Fig.~\ref{fig:fig4} for the various observables at the critical strain $\eta_{\rm c}$ follow the same format of Fig.~\ref{fig:fig2}, except that the hydrostatic tension $-p(\kappa)$ --- which is the relevant stress component under dilation --- is plotted in panel (c) instead of the shear stress. The numerical results are in excellent quantitative agreement with the predictions in Eqs.~(\ref{eq:displacement})-(\ref{eq:na_displacements_squared}), for both values of $\delta{z}$ used.
\begin{figure}[ht!]
  \includegraphics[width = 0.5\textwidth]{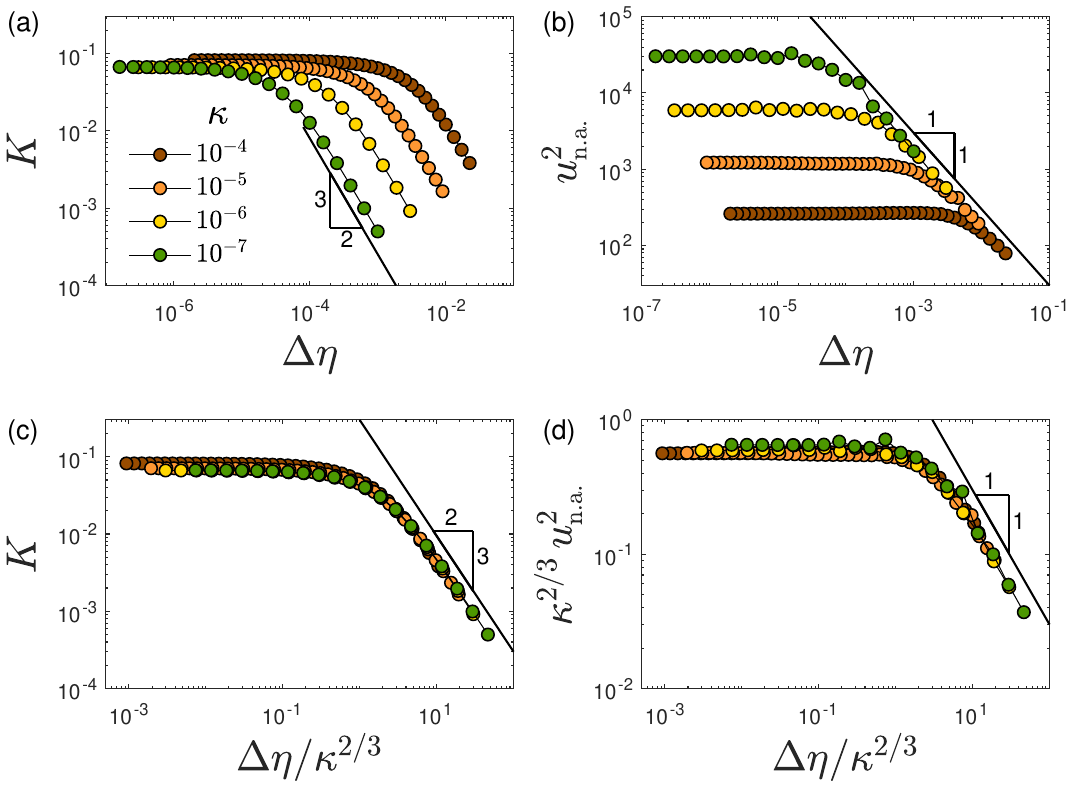}
  \caption{\small {\bf Numerical validation of the characteristic strain scale near the critical strain and of the strain-stiffening power-law scaling relations away from criticality under dilation.} The same as Fig.~\ref{fig:fig3}, but for dilation. Here, the distance from the critical point corresponds to $\Delta\eta\!=\!\eta_{\rm c}-\eta$ and the differential bulk modulus $K(\Delta\eta,\kappa)$ is plotted in panels (a) and (c). Excellent agreement with the theoretical predictions in Eqs.~(\ref{eq:na_displacements_squared})-(\ref{eq:away_from_critical_b}) is demonstrated.}
  \label{fig:fig5}
\end{figure}

The results presented in Fig.~\ref{fig:fig4} provide strong support for the universality of the strain-stiffening transition with respect to the network's topology and the symmetry of the driving strain. If this is the case, then the predictions in Eqs.~(\ref{eq:typical_strain})-(\ref{eq:away_from_critical_b}) should equally hold when applied to dilatational straining involving the rescaled variable $\Delta\eta/\delta\eta_*(\kappa)\!\equiv\!(\eta_{\rm c}-\eta)/\delta\eta_*(\kappa)$. This is indeed demonstrated in Fig.~\ref{fig:fig5}, which remarkably mirrors the simple shear results of Fig.~\ref{fig:fig3}, where the differential bulk modulus $K$ is used for ${\cal K}$ instead of $G$. The results presented in Fig.~\ref{fig:fig5} for a single value of $\delta{z}\!=\!0.5$ are also shown in the {\color{blue}{\it Appendix}} to be valid for another $\delta{z}$ value. Overall, the numerical results presented above provide strong support to the analytic predictions in Eqs.~(\ref{eq:displacement})-(\ref{eq:away_from_critical_b}), also demonstrating the independence of all scaling relations and critical exponents of the symmetry of the applied strain and the network's topology. Additional supporting results in relation to quantities not included in Eqs.~(\ref{eq:displacement})-(\ref{eq:away_from_critical_b}) are provided in Sect.~S-2 of the {\color{blue}{\it Appendix}}. Our findings thus strongly support the recently developed theory of the strain-stiffening transition in athermal disordered networks.

\section{D\lowercase{iscussion}}

In this work, we provided compelling numerical support to a comprehensive theory of the strain-stiffening phase transition in polymer networks, including all scaling relations and critical exponents in Eqs.~(\ref{eq:displacement})-(\ref{eq:away_from_critical_b}), and additional physical quantities discussed in the {\color{blue}{\it Appendix}}. In addition, the universality of the theoretical predictions with respect to the network’s topology, as quantified by the degree of connectivity $\delta{z}$ below the Maxwell threshold, and the symmetry of the driving strain --- i.e., shear and dilatational straining --- has been demonstrated.

The conceptual decomposition of the system into interacting stiff and soft sub-networks and the accompanying two-step straining procedure, which underlay the theory, are also useful in the context of the computer simulations as they allow to determine the critical $\epsilon_{\rm c}$ ambiguously and accurately --- which is difficult to do otherwise --- and make the computations more efficient as energy minimization is substantially less demanding in the absence of weak bending interactions upon forward-straining (since the energy of the stiff sub-system vanishes identically for $\epsilon\!<\!\epsilon_{\rm c}$). Future work should extend the present 2D numerical simulations to three dimensions (3D).

Future work should also address the possible existence of a $\kappa$-dependent lengthscale $\xi(\kappa)$ associated with the transition, possibly being divergent in the $\kappa\!\to\!0$ limit, as shown for isostatic isotropic spring networks in~\cite{lerner2024effects}. The possible existence of another characteristic strain scale above the critical point should be clarified as well. Moreover, while we established the $\delta{z}$-independence of the critical exponents, it would be interesting to clarify the dependence of the pre-factors on $\delta{z}$. Recent work in this direction indicates that the scaling laws tested here are valid in the regime $\kappa/\delta{z}^2\!\ll\!1$~\cite{lerner2024effects}.

Finally, while experiments fall short of quantitatively probing many of the theoretically-predicted quantities, and even the experimental determination of the critical strain $\epsilon_{\rm c}$ is not direct, it would be interesting to systematically analyze experimental data for the differential modulus ${\cal K}(\epsilon,\kappa)$ --- similar to the one shown in Fig.~\ref{fig:fig1} --- in light of the prediction in Eq.~\eqref{eq:away_from_critical_a}. Attempts to test the predicted universality of the strain-stiffening transition with respect to the symmetry of the driving strain would be also of interest.

\acknowledgements

E.B.~acknowledges support from the Israel Science Foundation (ISF Grant No.~403/24),  the Ben May Center for Chemical Theory and Computation and the Harold Perlman Family.

\clearpage

\appendix




\setcounter{figure}{0}
\renewcommand{\thefigure}{A\arabic{figure}}

\section{Computer simulations}
\label{sec:math_simulations}

In this section, we provide technical details regarding the computer simulations used to test the theoretical predictions. In Sect.~\ref{subsec:symmetry_moduli}, the symmetry of the driving strain and the differential elastic moduli are discussed. In Sect.~\ref{subsec:interactions_two_step}, the sub-network decomposition and the two-step straining procedure are discussed. 

\subsection{The initial isotropic networks}
\label{subsec:isotropic_networks}

To generate disordered computer networks, we first create packings of harmonic discs --- similarly to the ones studied, e.g.,~in~\cite{breakdown} ---, at a packing fraction of 1.0. We then adopt the network of contacts between the harmonic discs to form an initial disordered network of nodes and edges, featuring $z\!>\!z_{\rm c}$. The edges of these initial random networks are then diluted to achieve a target $z\!<\!z_{\rm c}$ following the algorithm described in~\cite{quantifier_PRE_2021}, which maintains low fluctuations in the local connectivity of nodes.

\subsection{The symmetry of the driving strain and the differential elastic moduli}
\label{subsec:symmetry_moduli}

As discussed in the main text, we tested the strain-stiffening theoretical predictions in large-scale numerical simulations for two qualitatively different straining symmetries. The first corresponds to simple shear straining, see Fig.~\ref{fig:figS1}a. It corresponds to a homogeneous (space independent) 2D deformation gradient tensor
\begin{equation}
\bm{F}_{\rm s}(\gamma) = \left( \begin{array}{cc}
1&\gamma\\
0&1
\end{array}\right)\ .
\label{eq:shear}
\end{equation}
The latter is parameterized by a strain amplitude $\gamma$ and is applied according to $\xv\!=\! \bm{F}_{\rm s}(\gamma)\!\cdot\!\Xv$ to a reference 2D network of an initial linear size $L$ (composed of $N$ nodes). Here, $\Xv$ is the coordinate describing the reference, unstrained network and $\xv$ describes the strained network (note that since $\bm{F}_{\rm s}(\gamma)$ is independent of $\Xv$, we also have $d\xv\!=\! \bm{F}_{\rm s}(\gamma)\!\cdot\! d\Xv$, which is the conventional definition of the deformation gradient tensor as relating vectorial line elements/edges in the unstrained and strained configurations).
\begin{figure}[htbp!]
    \centering
    \includegraphics[width=1\linewidth]{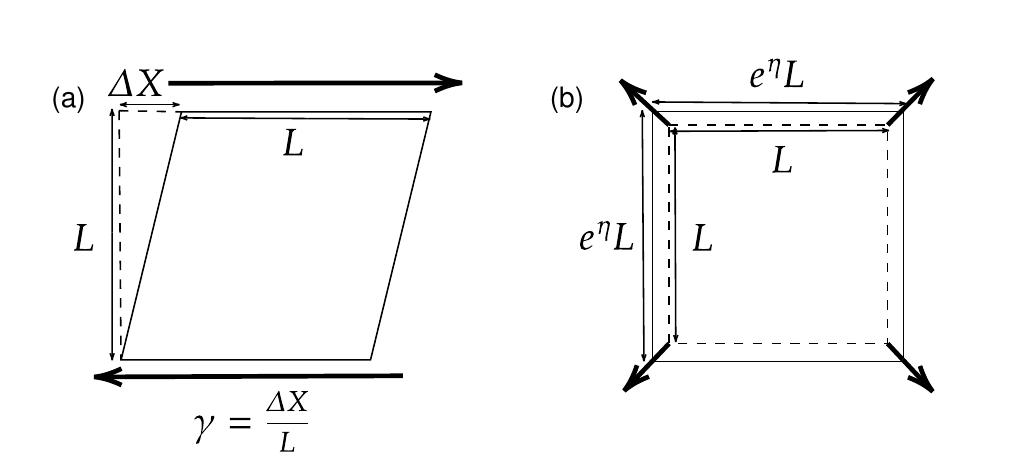}
	\caption{\textbf{Two symmetries of the driving strain.} (a) A simple shear strain applied to a 2D network of an initial linear size $L$. (b) A bulk expansion/dilation applied to the same network. See text for details and discussion.}
    \label{fig:figS1}
\end{figure}

Denoting the energy of the network by $U$, the shear stress is defined as
\begin{equation}
\sigma \equiv  \frac{1}{V}\frac{{\cal D}U}{{\cal D}\gamma} \ ,
\end{equation}
where $V$ is the volume of the network and the operator ${\cal D}/{\cal D\gamma}$ denotes a derivative with respect to $\gamma$ under mechanical equilibrium conditions. The differential shear modulus is defined as
\begin{equation}
G \equiv \frac{{\cal D}\sigma}{{\cal D}\gamma}  = \frac{1}{V}\frac{{\cal D}^2U}{{\cal D}\gamma^2} \ ,
\label{eq:G}
\end{equation}
where we used the fact that the volume $V$ is fixed under simple shear straining.

While the homogeneous deformation gradient tensor $\bm{F}_{\rm s}(\gamma)$ corresponds to an affine deformation, the intrinsic disorder of the network implies that the actual network's deformation contains a non-affine contribution. This is manifested in various physical observables, including $G$ that takes the form~\cite{lutsko1989generalized}
\begin{align}
   G = G_{\mbox{\tiny Born}} - \frac{1}{V}\left(\mathbfcal{F}_{\gamma}\cdot\mathbfcal{H}^{-1}\cdot \mathbfcal{F}_{\gamma}\right)\,,
    \label{eq:G_Born_nonaffine}
\end{align}
where $G_{\mbox{\tiny Born}}\!\equiv\!\frac{1}{V}\frac{\partial^2 U}{\partial \gamma^2}$ corresponds to the affine contribution and the second term corresponds to the non-affine contribution. In the latter, $\mathbfcal{H}^{-1}$ is the inverse of the Hessian matrix $\mathbfcal{H}\! \equiv \! \frac{\partial^2 U}{\partial \xv \partial \xv}$ and $\mathbfcal{F}_{\gamma}\!\equiv\!-\frac{\partial^2U}{\partial \gamma \partial \xv}$ is the non-affine force. Finally, we define the non-affine displacements as
\begin{equation}
 {\bm u}_{\mbox{\tiny{n.a.}}} \equiv -\mathbfcal{H}^{-1}\cdot \mathbfcal{F}_{\gamma} \ .
 \label{eq:non-affine_displ_vec}
\end{equation}
The non-affine displacements squared $u_{\mbox{\tiny n.a.}}^2$, discussed in the main text, are then defined as
\begin{equation}
    u_{\mbox{\tiny n.a.}}^2 \equiv \frac{1}{N} \sum_{i=1}^N {\bm u}^{(i)}_{\mbox{\tiny{n.a.}}} \cdot {\bm u}^{(i)}_{\mbox{\tiny{n.a.}}} \ ,
    \label{eq:non-affine_displ_squared}
\end{equation}
where $i$ is a nodal index and ${\bm u}^{(i)}_{\mbox{\tiny{n.a.}}}$ corresponds to the Cartesian non-affine displacement vector associated with node $i$.

The second straining symmetry we consider corresponds to bulk expansion/dilation, see Fig.~\ref{fig:figS1}b. In this case, the 2D deformation gradient tensor takes the form
\begin{equation}
\bm{F}_{\rm d}(\eta) = \left(\begin{array}{cc}e^\eta&0\\0&e^\eta\end{array}\right)\,,
\label{eq:dilation}
\end{equation}
which is parameterized by the strain amplitude $\eta$, corresponding to the logarithmic (Hencky) strain (see below). An obvious and important difference compared to the simple shear case is that the volume of the network is no longer constant during straining. Denoting the volume of the unstrained network by $\Omega\=L^\dbar$, we have $V\=e^{\eta\dbar}\Omega$ (according to Eq.~\eqref{eq:dilation}) for the volume of the strained network, where $\dbar$ is space dimension (we keep $\dbar$ in the expressions below, while $\bm{F}_{\rm d}(\eta)$ in Eq.~\eqref{eq:dilation} and the actual numerical simulations are 2D, i.e., $\dbar\=2$). These relations indeed indicate that $\eta\=\log\big(V^{1/\dbar}/\Omega^{1/\dbar}\big)\=\sFrac{1}{\dbar}\log\big(V/\Omega\big)$, i.e., that $\eta$ is the logarithmic (Hencky) strain. Note that under simple shear, which is volume preserving, we had $V\=\Omega$ throughout the straining process.

The relevant stress component under dilation is the pressure, given by
\begin{equation}
p \equiv -\frac{{\cal D} U}{{\cal D} V} = -\frac{{\cal D} U}{{\cal D} \eta}\frac{\partial \eta}{\partial V} = -\frac{1}{\dbar V}\frac{{\cal D} U}{{\cal D}\eta}\,.
\end{equation}
Note that in the main text we report the hydrostatic tension $-p$. The differential bulk modulus is then defined as
\begin{equation}
    K \equiv -V\frac{{\cal D} p}{{\cal D}V} = \frac{1}{\dbar^2 V}\frac{{\cal D}^2 U}{{\cal D} \eta^2} + p\,.
    \label{eq:K_Born_nonaffine}
\end{equation}
Similarly to Eq.~\eqref{eq:G_Born_nonaffine}, we have
\begin{align}
    K = K_{\mbox{\tiny Born}} - \frac{1}{\dbar^2 V}\left(\mathbfcal{F}_{\eta}\cdot\mathbfcal{H}^{-1}\cdot \mathbfcal{F}_{\eta}\right) + p\,,
    \label{bulk_modulus_equation}
\end{align}
where $K_{\mbox{\tiny Born}}\!\equiv\!\frac{1}{\dbar^2 V}\frac{\partial^2 U}{\partial \eta^2}$ and $\mathbfcal{F}_{\eta}\!\equiv\!-\frac{\partial^2U}{\partial \eta \partial \xv}$. The non-affine displacements are defined similarly to Eqs.~\eqref{eq:non-affine_displ_vec}-\eqref{eq:non-affine_displ_squared}.

\subsection{Sub-network decomposition and the two-step straining procedure}
\label{subsec:interactions_two_step}

As explained in the main text, we decompose the total energy of the network into a stiff sub-network contribution $U_{\mbox{\tiny stiff}}$, emerging from stretching/compression of the network's bonds/edges, and a soft (bending) sub-network contribution $U_{\mbox{\tiny soft}}$. For the former, we have
\begin{equation}
    U_{\mbox{\tiny stiff}}(\xv) = \frac{1}{2}\,\mu\!\!\!\sum_{\mbox{\tiny edges }ij}(r_{ij} - l_{ij})^2\ ,
    \label{eq:stiff_interaction}
\end{equation}
where $ij$ corresponds to each pair of interacting nodes, forming a bond (a network's edge). $l_{ij}$ is the rest-length of the bond and $r_{ij}$ is the deformed length, obtained from the nodal positions $\xv$ of the strained network. Here, $\mu$ is the stretching/compression stiffness of the bonds.

As explained in the main text, the stiff sub-network is forward-strained to the critical point $\epsilon_{\rm c}$. Then, the soft sub-network is added and a new equilibrium state is found through energy minimization (see details below). The soft sub-network features the following bending interaction energy
\begin{align}
    U_{\mbox{\tiny soft}}(\xv) = \frac{1}{2}\,\kappa_\theta\!\!\!\!\! \sum_{\mbox{\tiny angles }ijk}\!\!\!\left[\theta_{ijk}-\theta^{(0)}_{ijk}\right]^2\ ,
    \label{eq:soft_interaction}
\end{align}
where $\kappa_\theta$ is the bending rigidity. Here, $\theta_{ijk}$ corresponds to the angle formed at network's node $j$, featuring bonds with two adjacent nodes $i$ and $k$. Likewise, $\theta^{(0)}_{ijk}$ is the corresponding rest-angle defined {\it at the undeformed state corresponding to $\epsilon\!=\!0$}. We normalized all lengths by $\ell\!\equiv\!(\Omega/N)^{1/\dbar}$ and set $\kappa\!\equiv\!\kappa_\theta \ell^2/\mu\!\ll\!1$, extensively used in the main text.

A characteristic displacement scale $u_\star$, defined at $\epsilon_{\rm c}$, is obtained as follows: for each network we compute the magnitude of the node-wise displacements $|\uv_\star^{(i)}|$ between the $\kappa\!=\!0$ state and the $\kappa\!>\!0$ state. The node-wise distribution of $|\uv_\star^{(i)}|$ is broad; in order to reliably extract a typical scale from this noisy observable, we compute the {\it median} of $|\uv_\star^{(i)}|$ over the network's $N$ nodes. That median is then averaged over all our networks, to obtain $u_\star$ reported in Figs.~\ref{fig:fig2}a and~\ref{fig:fig4}a of the main text.

Physical observables computed away from the critical strain, i.e., at strains $\epsilon\!<\!\epsilon_{\rm c}$, are obtained by back-straining the network (with the two interacting sub-networks) from $\epsilon_{\rm c}$ to a desired strain $\epsilon$ in small strain steps, minimizing the energy in each step. Throughout the athermal and quasistatic straining process, Lees-Edwards boundary conditions are employed~\cite{rens2019theory}. Energy minimization in each strain step is performed using the FIRE minimization technique~\cite{bitzek2006structural}. The minimization process is terminated either when the typical compressive/tensile forces in the stiff sub-network's edges drop below $10^{-10}\mu\ell$ (for networks below the critical strain), or when the ratio of the typical net force on the network nodes, to the typical compressive/tensile forces in the stiff sub-network edges, drops below $10^{-8}$.


\setcounter{figure}{0}
\renewcommand{\thefigure}{B\arabic{figure}}

\section{A\lowercase{dditional supporting results}}
\label{sec:additional_results}

In this section, we aim at proving additional supporting results, either for quantities discussed in the main text or for other quantities not discussed therein, but for which theoretical predictions are available.
\begin{figure*}[ht!]
 \includegraphics[width = 0.8\textwidth]{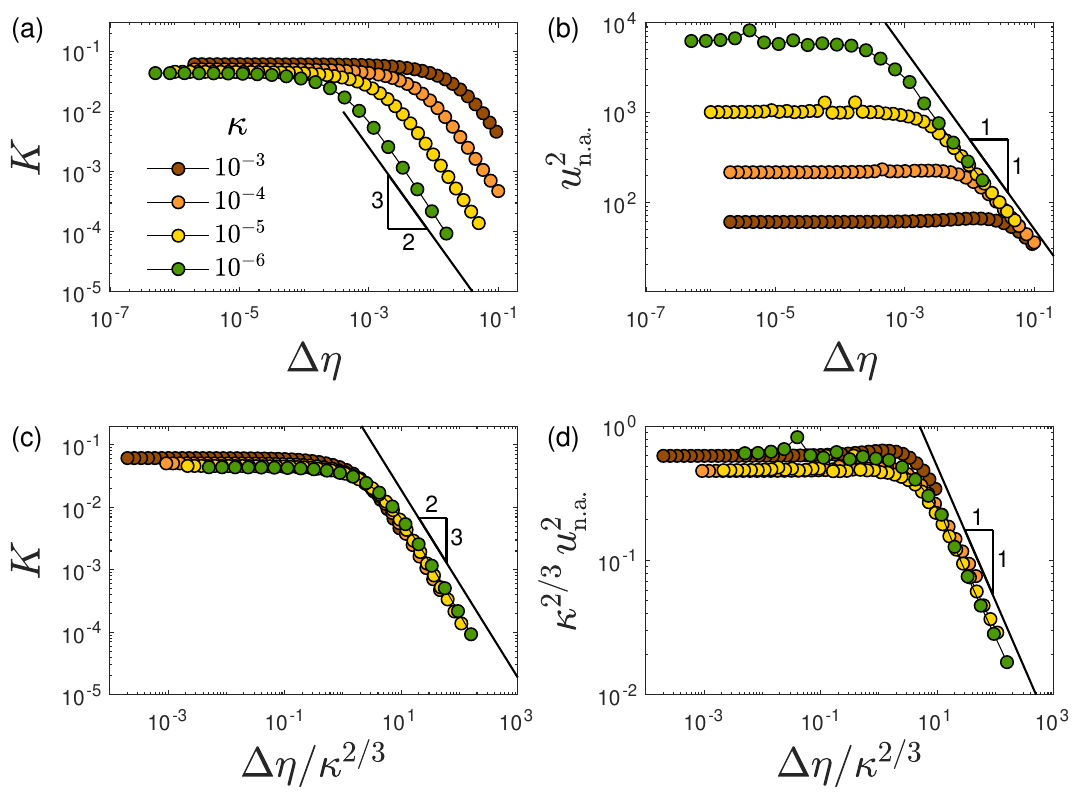}
  \caption{\small The same as Fig.~\ref{fig:fig5} in the main text (where $\delta{z}\!=\!0.5$), but for $\delta{z}\!=\!1$. Excellent
agreement with the theoretical predictions in Eqs.~(4)-(7) is demonstrated, which together with Fig.~\ref{fig:fig4} in the main text, provide strong support to the universality of the theory with respect to $\delta{z}$.}
  \label{fig:figS2}
\end{figure*}

\subsection{Additional support to the universality with respect to $\delta{z}$}
\label{sec:delta_z}

In Fig.~\ref{fig:fig5} in the main text, we provide numerical validation of the characteristic strain scale near the critical strain and of the strain-stiffening power-law scaling relations away from criticality under dilation for $\delta{z}\=0.5$ (i.e., $z\=3.5$). In Fig.~\ref{fig:figS2}, we demonstrate that these theoretical predictions remain valid as $\delta{z}$ is varied, which in conjunction with Fig.~~\ref{fig:fig4} in the main text, provide strong support to the universality of the theory with respect to $\delta{z}$.

\subsection{Numerical validation of the prediction $d{\cal K}(\epsilon_{\rm c},\kappa)/d\epsilon\!\sim\!\kappa^{-2/3}$}
\label{sec:derivative_modulus_at_criticality}

In the main text, the theoretical prediction~\cite{lerner2023scaling}
\begin{equation}
 \frac{d{\cal K}(\epsilon_{\rm c},\kappa)}{d\epsilon}\sim\kappa^{-2/3}
 \label{eq:derivative_modulus_at_criticality}
\end{equation}
is stated. In Fig.~\ref{fig:figS3}, we provide numerical validation of this prediction for both shear and dilation, i.e., for $dG(\gamma_{\rm c},\kappa)/d\gamma$ and $dK(\eta_{\rm c},\kappa)/d\eta$.

\begin{figure}[ht!]
 \includegraphics[width = 0.5\textwidth]{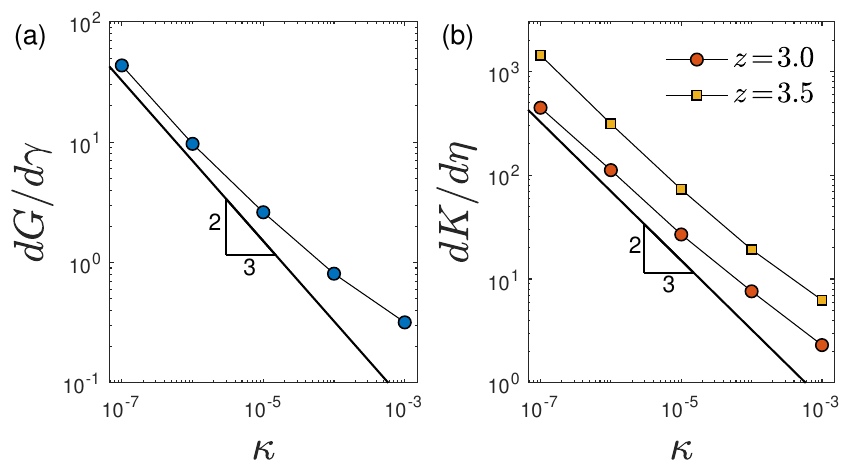}
  \caption{\small Numerical validation of the theoretical prediction in Eq.~\eqref{eq:derivative_modulus_at_criticality}, for both shear (panel (a), $dG(\gamma_{\rm c},\kappa)/d\gamma$ for $z\!=\!3.5$) and dilation (panel (b), $dK(\eta_{\rm c},\kappa)/d\eta$ for $z\!=\!3.0, 3.5$, see legend).}
  \label{fig:figS3}
\end{figure}

\subsection{T\lowercase{he geometric operator defining states-of-self-stress and the associated eigenvalue}}
\label{sec:SSS_eigenvalue}

A basic object characterizing the strain-stiffening transition for $\kappa\=0$, not discussed in the main text, is the state-of-self-stress (SSS). A SSS corresponds to a set of putative bond forces of magnitude $f_{ij}$ that exactly balance each other on every node in the network~\cite{Calladine_1978}, i.e., $\sum_{\mbox{\tiny neighbors }j(i)} \nv_{ji}f_{ij}\=\zerovector$.
Here, $n_{ji}$ is a unit vector pointing from node $j$ to node $i$ and the right-hand-side corresponds to vanishing nodal resultant forces. It is convenient to write the above relation in the form $\Svv^T\ket{f}\=\ket{\zerovector}$, where $\Svv^T$ is an operator accounting for the network's geometry and $\ket{f}$ corresponding to the scalar bond forces.
Consider then the operator $\Svv\Svv^T$, featuring an eigenvalue $\lambda$, i.e., $\Svv\Svv^T\ket{f}\= \lambda\ket{f}$.
In the absence of soft bending interactions (corresponding to $\kappa\=0$), strain-stiffened
networks feature a single SSS at the critical strain $\epsilon_{\rm c}$, corresponding to $\lambda\=0$~\cite{during2014length,rens2018micromechanical}. The introduction of the soft sub-network with $\kappa\!>\!0$ at $\epsilon_{\rm c}\=0$ gives rise to a finite $\lambda\!\sim\!u_*^2$~\cite{lerner2023scaling}. Equation~(1) in the main text then implies~\cite{lerner2023scaling}
\begin{equation}
    \lambda(\epsilon_{\rm c},\kappa) \sim \kappa^{2/3}  \ .
    \label{eq:SSS_eigenvalue}
\end{equation}
In Fig.~\ref{fig:figS4}, we a provide numerical validation of this prediction for both shear and dilation, i.e., for $\lambda(\gamma_{\rm c},\kappa)$ and $\lambda(\eta_{\rm c},\kappa)$, and two values of $\delta{z}$.
\begin{figure}[ht!]
 \includegraphics[width = 0.45\textwidth]{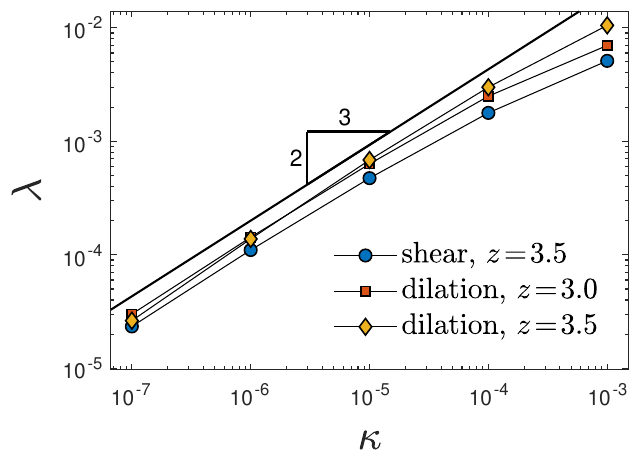}
  \caption{\small Numerical validation of the theoretical prediction in Eq.~\eqref{eq:SSS_eigenvalue}, for both shear ($\epsilon_{\rm c}\!=\!\gamma_{\rm c})$ and dilation ($\epsilon_{\rm c}\!=\!\eta_{\rm c}$), and two values of $z$ (see legend)}.
  \label{fig:figS4}
\end{figure}

\subsection{The differential modulus at criticality and its singular perturbation nature}
\label{sec:jump_discontinuity}

\begin{figure}[ht!]
 \includegraphics[width = 0.5\textwidth]{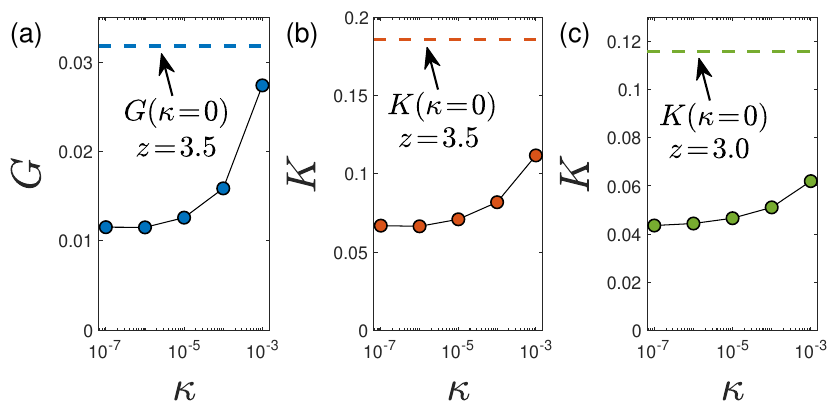}
  \caption{\small ${\cal K}(\epsilon_{\rm c},\kappa)$ vs.~$\kappa\!>\!0$ (circles) under shear straining and $z\!=\!3.5$ (panel (a)), under dilatational straining and $z\!=\!3.5$ (panel (b)) and $z\!=\!3.0$  (panel (c)). The corresponding value for ${\cal K}(\epsilon_{\rm c},\kappa\!=\!0)$ are shown by the horizontal dashed lines.}
  \label{fig:figS5}
\end{figure}

In the main text, it is stated that for $\kappa\=0$, the differential modulus features a jump discontinuity at critically, attaining a finite value ${\cal K}(\epsilon_{\rm c},\kappa\=0)$.

Moreover, it is stated therein that for $\kappa\!>\!0$ and for $\epsilon$ sufficiently close to $\epsilon_{\rm c}$, one has ${\cal K}(\epsilon,\kappa)-{\cal K}(\epsilon_{\rm c})\!\sim\!\kappa^{-2/3}(\epsilon-\epsilon_{\rm c})$. Here, ${\cal K}(\epsilon_{\rm c})$ corresponds to ${\cal K}(\epsilon_{\rm c}, \kappa\!\to\!0^+)$. It was shown in~\cite{lerner2023scaling} that in fact ${\cal K}(\epsilon_{\rm c},\kappa\!\to\!0^+)$ differs from ${\cal K}(\epsilon_{\rm c},\kappa\=0)$ and satisfies
\begin{equation}
    {\cal K}(\epsilon_{\rm c},\kappa\!\to\!0^+)={\cal K}(\epsilon_{\rm c},\kappa\=0) -\Delta{\cal K} \equiv {\cal K}(\epsilon_{\rm c}) \ ,
    \label{eq:jump_discontinuity}
\end{equation}
with $\Delta{\cal K}\!>\!0$. The existence of a finite $\Delta{\cal K}$ is yet another manifestation of the singular perturbation nature of the soft/weak bending interactions associated with $\kappa\!>\!0$.

Here, we provide numerical support to the prediction in Eq.~\eqref{eq:jump_discontinuity} for both shear and dilation. This is achieved by plotting
$G(\gamma_{\rm c},\kappa\=0)$ vs.~$G(\gamma_{\rm c},\kappa\!>\!0)$ for $z\=3.5$ in Fig.~\ref{fig:figS5}a, and $K(\eta_{\rm c},\kappa\=0)$ vs.~$K(\eta_{\rm c},\kappa\!>\!0)$ for $z\=3.5$ in Fig.~\ref{fig:figS5}b and for $z\=3.0$ in Fig.~\ref{fig:figS5}c. In the $\kappa\!\to\!0^+$ limit, the data in the three panels appear to be consistent with ${\cal K}(\epsilon_{\rm c},\kappa\!\to\!0^+)/{\cal K}(\epsilon_{\rm c},\kappa\=0)\=1/3$, suggested in~\cite{freddy_mac_prl_2024}.


\begin{thebibliography}{38}%
\makeatletter
\providecommand \@ifxundefined [1]{%
 \@ifx{#1\undefined}
}%
\providecommand \@ifnum [1]{%
 \ifnum #1\expandafter \@firstoftwo
 \else \expandafter \@secondoftwo
 \fi
}%
\providecommand \@ifx [1]{%
 \ifx #1\expandafter \@firstoftwo
 \else \expandafter \@secondoftwo
 \fi
}%
\providecommand \natexlab [1]{#1}%
\providecommand \enquote  [1]{``#1''}%
\providecommand \bibnamefont  [1]{#1}%
\providecommand \bibfnamefont [1]{#1}%
\providecommand \citenamefont [1]{#1}%
\providecommand \href@noop [0]{\@secondoftwo}%
\providecommand \href [0]{\begingroup \@sanitize@url \@href}%
\providecommand \@href[1]{\@@startlink{#1}\@@href}%
\providecommand \@@href[1]{\endgroup#1\@@endlink}%
\providecommand \@sanitize@url [0]{\catcode `\\12\catcode `\$12\catcode
  `\&12\catcode `\#12\catcode `\^12\catcode `\_12\catcode `\%12\relax}%
\providecommand \@@startlink[1]{}%
\providecommand \@@endlink[0]{}%
\providecommand \url  [0]{\begingroup\@sanitize@url \@url }%
\providecommand \@url [1]{\endgroup\@href {#1}{\urlprefix }}%
\providecommand \urlprefix  [0]{URL }%
\providecommand \Eprint [0]{\href }%
\providecommand \doibase [0]{https://doi.org/}%
\providecommand \selectlanguage [0]{\@gobble}%
\providecommand \bibinfo  [0]{\@secondoftwo}%
\providecommand \bibfield  [0]{\@secondoftwo}%
\providecommand \translation [1]{[#1]}%
\providecommand \BibitemOpen [0]{}%
\providecommand \bibitemStop [0]{}%
\providecommand \bibitemNoStop [0]{.\EOS\space}%
\providecommand \EOS [0]{\spacefactor3000\relax}%
\providecommand \BibitemShut  [1]{\csname bibitem#1\endcsname}%
\let\auto@bib@innerbib\@empty
\bibitem [{\citenamefont {Picu}(2022)}]{picu2022network}%
  \BibitemOpen
  \bibfield  {author} {\bibinfo {author} {\bibfnamefont {C.~R.}\ \bibnamefont
  {Picu}},\ }\href@noop {} {\emph {\bibinfo {title} {Network materials:
  structure and properties}}}\ (\bibinfo  {publisher} {Cambridge University
  Press},\ \bibinfo {year} {2022})\BibitemShut {NoStop}%
\bibitem [{\citenamefont {Storm}\ \emph {et~al.}(2005)\citenamefont {Storm},
  \citenamefont {Pastore}, \citenamefont {MacKintosh}, \citenamefont
  {Lubensky},\ and\ \citenamefont {Janmey}}]{kees_nature_2005}%
  \BibitemOpen
  \bibfield  {author} {\bibinfo {author} {\bibfnamefont {C.}~\bibnamefont
  {Storm}}, \bibinfo {author} {\bibfnamefont {J.~J.}\ \bibnamefont {Pastore}},
  \bibinfo {author} {\bibfnamefont {F.~C.}\ \bibnamefont {MacKintosh}},
  \bibinfo {author} {\bibfnamefont {T.~C.}\ \bibnamefont {Lubensky}},\ and\
  \bibinfo {author} {\bibfnamefont {P.~A.}\ \bibnamefont {Janmey}},\ }\bibfield
   {title} {\bibinfo {title} {Nonlinear elasticity in biological gels},\ }\href
  {https://doi.org/10.1038/nature03521} {\bibfield  {journal} {\bibinfo
  {journal} {Nature}\ }\textbf {\bibinfo {volume} {435}},\ \bibinfo {pages}
  {191} (\bibinfo {year} {2005})}\BibitemShut {NoStop}%
\bibitem [{\citenamefont {Erk}\ \emph {et~al.}(2010)\citenamefont {Erk},
  \citenamefont {Henderson},\ and\ \citenamefont {Shull}}]{erk2010strain}%
  \BibitemOpen
  \bibfield  {author} {\bibinfo {author} {\bibfnamefont {K.~A.}\ \bibnamefont
  {Erk}}, \bibinfo {author} {\bibfnamefont {K.~J.}\ \bibnamefont {Henderson}},\
  and\ \bibinfo {author} {\bibfnamefont {K.~R.}\ \bibnamefont {Shull}},\
  }\bibfield  {title} {\bibinfo {title} {Strain stiffening in synthetic and
  biopolymer networks},\ }\href {https://doi.org/10.1021/bm100136y} {\bibfield
  {journal} {\bibinfo  {journal} {Biomacromolecules}\ }\textbf {\bibinfo
  {volume} {11}},\ \bibinfo {pages} {1358} (\bibinfo {year}
  {2010})}\BibitemShut {NoStop}%
\bibitem [{\citenamefont {Sharma}\ \emph
  {et~al.}(2016{\natexlab{a}})\citenamefont {Sharma}, \citenamefont {Licup},
  \citenamefont {Jansen}, \citenamefont {Rens}, \citenamefont {Sheinman},
  \citenamefont {Koenderink},\ and\ \citenamefont
  {MacKintosh}}]{robbie_nature_physics_2016}%
  \BibitemOpen
  \bibfield  {author} {\bibinfo {author} {\bibfnamefont {A.}~\bibnamefont
  {Sharma}}, \bibinfo {author} {\bibfnamefont {A.~J.}\ \bibnamefont {Licup}},
  \bibinfo {author} {\bibfnamefont {K.~A.}\ \bibnamefont {Jansen}}, \bibinfo
  {author} {\bibfnamefont {R.}~\bibnamefont {Rens}}, \bibinfo {author}
  {\bibfnamefont {M.}~\bibnamefont {Sheinman}}, \bibinfo {author}
  {\bibfnamefont {G.~H.}\ \bibnamefont {Koenderink}},\ and\ \bibinfo {author}
  {\bibfnamefont {F.~C.}\ \bibnamefont {MacKintosh}},\ }\bibfield  {title}
  {\bibinfo {title} {Strain-controlled criticality governs the nonlinear
  mechanics of fibre networks},\ }\href {https://doi.org/10.1038/nphys3628}
  {\bibfield  {journal} {\bibinfo  {journal} {Nature Physics}\ }\textbf
  {\bibinfo {volume} {12}},\ \bibinfo {pages} {584} (\bibinfo {year}
  {2016}{\natexlab{a}})}\BibitemShut {NoStop}%
\bibitem [{\citenamefont {Vatankhah-Varnosfaderani}\ \emph
  {et~al.}(2018)\citenamefont {Vatankhah-Varnosfaderani}, \citenamefont
  {Keith}, \citenamefont {Cong}, \citenamefont {Liang}, \citenamefont
  {Rosenthal}, \citenamefont {Sztucki}, \citenamefont {Clair}, \citenamefont
  {Magonov}, \citenamefont {Ivanov}, \citenamefont {Dobrynin},\ and\
  \citenamefont {Sheiko}}]{vatankhah2018chameleon}%
  \BibitemOpen
  \bibfield  {author} {\bibinfo {author} {\bibfnamefont {M.}~\bibnamefont
  {Vatankhah-Varnosfaderani}}, \bibinfo {author} {\bibfnamefont {A.~N.}\
  \bibnamefont {Keith}}, \bibinfo {author} {\bibfnamefont {Y.}~\bibnamefont
  {Cong}}, \bibinfo {author} {\bibfnamefont {H.}~\bibnamefont {Liang}},
  \bibinfo {author} {\bibfnamefont {M.}~\bibnamefont {Rosenthal}}, \bibinfo
  {author} {\bibfnamefont {M.}~\bibnamefont {Sztucki}}, \bibinfo {author}
  {\bibfnamefont {C.}~\bibnamefont {Clair}}, \bibinfo {author} {\bibfnamefont
  {S.}~\bibnamefont {Magonov}}, \bibinfo {author} {\bibfnamefont {D.~A.}\
  \bibnamefont {Ivanov}}, \bibinfo {author} {\bibfnamefont {A.~V.}\
  \bibnamefont {Dobrynin}},\ and\ \bibinfo {author} {\bibfnamefont {S.~S.}\
  \bibnamefont {Sheiko}},\ }\bibfield  {title} {\bibinfo {title}
  {Chameleon-like elastomers with molecularly encoded strain-adaptive
  stiffening and coloration},\ }\href {https://doi.org/10.1126/science.aar5308}
  {\bibfield  {journal} {\bibinfo  {journal} {Science}\ }\textbf {\bibinfo
  {volume} {359}},\ \bibinfo {pages} {1509} (\bibinfo {year}
  {2018})}\BibitemShut {NoStop}%
\bibitem [{\citenamefont {Levental}\ \emph {et~al.}(2007)\citenamefont
  {Levental}, \citenamefont {Georges},\ and\ \citenamefont
  {Janmey}}]{Janmey_soft_matter_2007}%
  \BibitemOpen
  \bibfield  {author} {\bibinfo {author} {\bibfnamefont {I.}~\bibnamefont
  {Levental}}, \bibinfo {author} {\bibfnamefont {P.~C.}\ \bibnamefont
  {Georges}},\ and\ \bibinfo {author} {\bibfnamefont {P.~A.}\ \bibnamefont
  {Janmey}},\ }\bibfield  {title} {\bibinfo {title} {Soft biological materials
  and their impact on cell function},\ }\href
  {https://doi.org/10.1039/B610522J} {\bibfield  {journal} {\bibinfo  {journal}
  {Soft Matter}\ }\textbf {\bibinfo {volume} {3}},\ \bibinfo {pages} {299}
  (\bibinfo {year} {2007})}\BibitemShut {NoStop}%
\bibitem [{\citenamefont {Winer}\ \emph {et~al.}(2009)\citenamefont {Winer},
  \citenamefont {Oake},\ and\ \citenamefont {Janmey}}]{winer2009non}%
  \BibitemOpen
  \bibfield  {author} {\bibinfo {author} {\bibfnamefont {J.~P.}\ \bibnamefont
  {Winer}}, \bibinfo {author} {\bibfnamefont {S.}~\bibnamefont {Oake}},\ and\
  \bibinfo {author} {\bibfnamefont {P.~A.}\ \bibnamefont {Janmey}},\ }\bibfield
   {title} {\bibinfo {title} {Non-linear elasticity of extracellular matrices
  enables contractile cells to communicate local position and orientation},\
  }\href {https://doi.org/10.1371/journal.pone.0006382} {\bibfield  {journal}
  {\bibinfo  {journal} {PloS one}\ }\textbf {\bibinfo {volume} {4}},\ \bibinfo
  {pages} {e6382} (\bibinfo {year} {2009})}\BibitemShut {NoStop}%
\bibitem [{\citenamefont {Das}\ \emph {et~al.}(2016)\citenamefont {Das},
  \citenamefont {Gocheva}, \citenamefont {Hammink}, \citenamefont {Zouani},\
  and\ \citenamefont {Rowan}}]{das2016stress}%
  \BibitemOpen
  \bibfield  {author} {\bibinfo {author} {\bibfnamefont {R.~K.}\ \bibnamefont
  {Das}}, \bibinfo {author} {\bibfnamefont {V.}~\bibnamefont {Gocheva}},
  \bibinfo {author} {\bibfnamefont {R.}~\bibnamefont {Hammink}}, \bibinfo
  {author} {\bibfnamefont {O.~F.}\ \bibnamefont {Zouani}},\ and\ \bibinfo
  {author} {\bibfnamefont {A.~E.}\ \bibnamefont {Rowan}},\ }\bibfield  {title}
  {\bibinfo {title} {Stress-stiffening-mediated stem-cell commitment switch in
  soft responsive hydrogels},\ }\href {https://doi.org/10.1038/nmat4483}
  {\bibfield  {journal} {\bibinfo  {journal} {Nature materials}\ }\textbf
  {\bibinfo {volume} {15}},\ \bibinfo {pages} {318} (\bibinfo {year}
  {2016})}\BibitemShut {NoStop}%
\bibitem [{\citenamefont {Han}\ \emph {et~al.}(2018)\citenamefont {Han},
  \citenamefont {Ronceray}, \citenamefont {Xu}, \citenamefont {Malandrino},
  \citenamefont {Kamm}, \citenamefont {Lenz}, \citenamefont {Broedersz},\ and\
  \citenamefont {Guo}}]{han2018cell}%
  \BibitemOpen
  \bibfield  {author} {\bibinfo {author} {\bibfnamefont {Y.~L.}\ \bibnamefont
  {Han}}, \bibinfo {author} {\bibfnamefont {P.}~\bibnamefont {Ronceray}},
  \bibinfo {author} {\bibfnamefont {G.}~\bibnamefont {Xu}}, \bibinfo {author}
  {\bibfnamefont {A.}~\bibnamefont {Malandrino}}, \bibinfo {author}
  {\bibfnamefont {R.~D.}\ \bibnamefont {Kamm}}, \bibinfo {author}
  {\bibfnamefont {M.}~\bibnamefont {Lenz}}, \bibinfo {author} {\bibfnamefont
  {C.~P.}\ \bibnamefont {Broedersz}},\ and\ \bibinfo {author} {\bibfnamefont
  {M.}~\bibnamefont {Guo}},\ }\bibfield  {title} {\bibinfo {title} {Cell
  contraction induces long-ranged stress stiffening in the extracellular
  matrix},\ }\href {https://doi.org/10.1073/pnas.1722619115} {\bibfield
  {journal} {\bibinfo  {journal} {Proc. Natl. Acad. Sci. U. S. A.}\ }\textbf
  {\bibinfo {volume} {115}},\ \bibinfo {pages} {4075} (\bibinfo {year}
  {2018})}\BibitemShut {NoStop}%
\bibitem [{\citenamefont {Wang}\ \emph {et~al.}(2023)\citenamefont {Wang},
  \citenamefont {Du},\ and\ \citenamefont {Xu}}]{wang2023strain}%
  \BibitemOpen
  \bibfield  {author} {\bibinfo {author} {\bibfnamefont {Y.}~\bibnamefont
  {Wang}}, \bibinfo {author} {\bibfnamefont {Y.}~\bibnamefont {Du}},\ and\
  \bibinfo {author} {\bibfnamefont {F.}~\bibnamefont {Xu}},\ }\bibfield
  {title} {\bibinfo {title} {Strain stiffening retards growth instability in
  residually stressed biological tissues},\ }\href
  {https://doi.org/10.1016/j.jmps.2023.105360} {\bibfield  {journal} {\bibinfo
  {journal} {J. Mech. Phys. Solids}\ }\textbf {\bibinfo {volume} {178}},\
  \bibinfo {pages} {105360} (\bibinfo {year} {2023})}\BibitemShut {NoStop}%
\bibitem [{\citenamefont {Maxwell}(1864)}]{maxwell_1864}%
  \BibitemOpen
  \bibfield  {author} {\bibinfo {author} {\bibfnamefont {J.~C.}\ \bibnamefont
  {Maxwell}},\ }\bibfield  {title} {\bibinfo {title} {L. on the calculation of
  the equilibrium and stiffness of frames},\ }\href
  {https://doi.org/10.1080/14786446408643668} {\bibfield  {journal} {\bibinfo
  {journal} {Philos. Mag.}\ }\textbf {\bibinfo {volume} {27}},\ \bibinfo
  {pages} {294} (\bibinfo {year} {1864})}\BibitemShut {NoStop}%
\bibitem [{\citenamefont {D{\"u}ring}\ \emph {et~al.}(2013)\citenamefont
  {D{\"u}ring}, \citenamefont {Lerner},\ and\ \citenamefont
  {Wyart}}]{phonon_gap_2012}%
  \BibitemOpen
  \bibfield  {author} {\bibinfo {author} {\bibfnamefont {G.}~\bibnamefont
  {D{\"u}ring}}, \bibinfo {author} {\bibfnamefont {E.}~\bibnamefont {Lerner}},\
  and\ \bibinfo {author} {\bibfnamefont {M.}~\bibnamefont {Wyart}},\ }\bibfield
   {title} {\bibinfo {title} {Phonon gap and localization lengths in floppy
  materials},\ }\href {https://doi.org/10.1039/C2SM25878A} {\bibfield
  {journal} {\bibinfo  {journal} {Soft Matter}\ }\textbf {\bibinfo {volume}
  {9}},\ \bibinfo {pages} {146} (\bibinfo {year} {2013})}\BibitemShut {NoStop}%
\bibitem [{\citenamefont {Licup}\ \emph {et~al.}(2015)\citenamefont {Licup},
  \citenamefont {M{\"u}nster}, \citenamefont {Sharma}, \citenamefont
  {Sheinman}, \citenamefont {Jawerth}, \citenamefont {Fabry}, \citenamefont
  {Weitz},\ and\ \citenamefont {MacKintosh}}]{licup2015stress}%
  \BibitemOpen
  \bibfield  {author} {\bibinfo {author} {\bibfnamefont {A.~J.}\ \bibnamefont
  {Licup}}, \bibinfo {author} {\bibfnamefont {S.}~\bibnamefont {M{\"u}nster}},
  \bibinfo {author} {\bibfnamefont {A.}~\bibnamefont {Sharma}}, \bibinfo
  {author} {\bibfnamefont {M.}~\bibnamefont {Sheinman}}, \bibinfo {author}
  {\bibfnamefont {L.~M.}\ \bibnamefont {Jawerth}}, \bibinfo {author}
  {\bibfnamefont {B.}~\bibnamefont {Fabry}}, \bibinfo {author} {\bibfnamefont
  {D.~A.}\ \bibnamefont {Weitz}},\ and\ \bibinfo {author} {\bibfnamefont
  {F.~C.}\ \bibnamefont {MacKintosh}},\ }\bibfield  {title} {\bibinfo {title}
  {Stress controls the mechanics of collagen networks},\ }\href
  {https://doi.org/10.1073/pnas.1504258112} {\bibfield  {journal} {\bibinfo
  {journal} {Proc. Natl. Acad. Sci. U. S. A.}\ }\textbf {\bibinfo {volume}
  {112}},\ \bibinfo {pages} {9573} (\bibinfo {year} {2015})}\BibitemShut
  {NoStop}%
\bibitem [{\citenamefont {Broedersz}\ and\ \citenamefont
  {MacKintosh}(2014)}]{RevModPhys.86.995}%
  \BibitemOpen
  \bibfield  {author} {\bibinfo {author} {\bibfnamefont {C.~P.}\ \bibnamefont
  {Broedersz}}\ and\ \bibinfo {author} {\bibfnamefont {F.~C.}\ \bibnamefont
  {MacKintosh}},\ }\bibfield  {title} {\bibinfo {title} {Modeling semiflexible
  polymer networks},\ }\href {https://doi.org/10.1103/RevModPhys.86.995}
  {\bibfield  {journal} {\bibinfo  {journal} {Rev. Mod. Phys.}\ }\textbf
  {\bibinfo {volume} {86}},\ \bibinfo {pages} {995} (\bibinfo {year}
  {2014})}\BibitemShut {NoStop}%
\bibitem [{\citenamefont {D\"uring}\ \emph
  {et~al.}(2014{\natexlab{a}})\citenamefont {D\"uring}, \citenamefont
  {Lerner},\ and\ \citenamefont {Wyart}}]{gustavo_pre_2014}%
  \BibitemOpen
  \bibfield  {author} {\bibinfo {author} {\bibfnamefont {G.}~\bibnamefont
  {D\"uring}}, \bibinfo {author} {\bibfnamefont {E.}~\bibnamefont {Lerner}},\
  and\ \bibinfo {author} {\bibfnamefont {M.}~\bibnamefont {Wyart}},\ }\bibfield
   {title} {\bibinfo {title} {Length scales and self-organization in dense
  suspension flows},\ }\href {https://doi.org/10.1103/PhysRevE.89.022305}
  {\bibfield  {journal} {\bibinfo  {journal} {Phys. Rev. E}\ }\textbf {\bibinfo
  {volume} {89}},\ \bibinfo {pages} {022305} (\bibinfo {year}
  {2014}{\natexlab{a}})}\BibitemShut {NoStop}%
\bibitem [{\citenamefont {Rens}\ \emph
  {et~al.}(2018{\natexlab{a}})\citenamefont {Rens}, \citenamefont {Villarroel},
  \citenamefont {D\"uring},\ and\ \citenamefont {Lerner}}]{robbie_pre_2018}%
  \BibitemOpen
  \bibfield  {author} {\bibinfo {author} {\bibfnamefont {R.}~\bibnamefont
  {Rens}}, \bibinfo {author} {\bibfnamefont {C.}~\bibnamefont {Villarroel}},
  \bibinfo {author} {\bibfnamefont {G.}~\bibnamefont {D\"uring}},\ and\
  \bibinfo {author} {\bibfnamefont {E.}~\bibnamefont {Lerner}},\ }\bibfield
  {title} {\bibinfo {title} {Micromechanical theory of strain stiffening of
  biopolymer networks},\ }\href {https://doi.org/10.1103/PhysRevE.98.062411}
  {\bibfield  {journal} {\bibinfo  {journal} {Phys. Rev. E}\ }\textbf {\bibinfo
  {volume} {98}},\ \bibinfo {pages} {062411} (\bibinfo {year}
  {2018}{\natexlab{a}})}\BibitemShut {NoStop}%
\bibitem [{\citenamefont {Merkel}\ \emph {et~al.}(2019)\citenamefont {Merkel},
  \citenamefont {Baumgarten}, \citenamefont {Tighe},\ and\ \citenamefont
  {Manning}}]{merkel_pnas_2019}%
  \BibitemOpen
  \bibfield  {author} {\bibinfo {author} {\bibfnamefont {M.}~\bibnamefont
  {Merkel}}, \bibinfo {author} {\bibfnamefont {K.}~\bibnamefont {Baumgarten}},
  \bibinfo {author} {\bibfnamefont {B.~P.}\ \bibnamefont {Tighe}},\ and\
  \bibinfo {author} {\bibfnamefont {M.~L.}\ \bibnamefont {Manning}},\
  }\bibfield  {title} {\bibinfo {title} {A minimal-length approach unifies
  rigidity in underconstrained materials},\ }\href
  {https://doi.org/10.1073/pnas.1815436116} {\bibfield  {journal} {\bibinfo
  {journal} {Proc. Natl. Acad. Sci. U.S.A.}\ }\textbf {\bibinfo {volume}
  {116}},\ \bibinfo {pages} {6560} (\bibinfo {year} {2019})}\BibitemShut
  {NoStop}%
\bibitem [{\citenamefont {Licup}\ \emph {et~al.}(2016)\citenamefont {Licup},
  \citenamefont {Sharma},\ and\ \citenamefont
  {MacKintosh}}]{mackintosh_pre_2016}%
  \BibitemOpen
  \bibfield  {author} {\bibinfo {author} {\bibfnamefont {A.~J.}\ \bibnamefont
  {Licup}}, \bibinfo {author} {\bibfnamefont {A.}~\bibnamefont {Sharma}},\ and\
  \bibinfo {author} {\bibfnamefont {F.~C.}\ \bibnamefont {MacKintosh}},\
  }\bibfield  {title} {\bibinfo {title} {Elastic regimes of subisostatic
  athermal fiber networks},\ }\href
  {https://doi.org/10.1103/PhysRevE.93.012407} {\bibfield  {journal} {\bibinfo
  {journal} {Phys. Rev. E}\ }\textbf {\bibinfo {volume} {93}},\ \bibinfo
  {pages} {012407} (\bibinfo {year} {2016})}\BibitemShut {NoStop}%
\bibitem [{\citenamefont {Rens}\ \emph {et~al.}(2016)\citenamefont {Rens},
  \citenamefont {Vahabi}, \citenamefont {Licup}, \citenamefont {MacKintosh},\
  and\ \citenamefont {Sharma}}]{Rens_JPCB_2016}%
  \BibitemOpen
  \bibfield  {author} {\bibinfo {author} {\bibfnamefont {R.}~\bibnamefont
  {Rens}}, \bibinfo {author} {\bibfnamefont {M.}~\bibnamefont {Vahabi}},
  \bibinfo {author} {\bibfnamefont {A.~J.}\ \bibnamefont {Licup}}, \bibinfo
  {author} {\bibfnamefont {F.~C.}\ \bibnamefont {MacKintosh}},\ and\ \bibinfo
  {author} {\bibfnamefont {A.}~\bibnamefont {Sharma}},\ }\bibfield  {title}
  {\bibinfo {title} {Nonlinear mechanics of athermal branched biopolymer
  networks},\ }\href {https://doi.org/10.1021/acs.jpcb.6b00259} {\bibfield
  {journal} {\bibinfo  {journal} {J. Phys. Chem. B}\ }\textbf {\bibinfo
  {volume} {120}},\ \bibinfo {pages} {5831} (\bibinfo {year}
  {2016})}\BibitemShut {NoStop}%
\bibitem [{\citenamefont {Sharma}\ \emph
  {et~al.}(2016{\natexlab{b}})\citenamefont {Sharma}, \citenamefont {Licup},
  \citenamefont {Rens}, \citenamefont {Vahabi}, \citenamefont {Jansen},
  \citenamefont {Koenderink},\ and\ \citenamefont
  {MacKintosh}}]{sharma2016strain}%
  \BibitemOpen
  \bibfield  {author} {\bibinfo {author} {\bibfnamefont {A.}~\bibnamefont
  {Sharma}}, \bibinfo {author} {\bibfnamefont {A.~J.}\ \bibnamefont {Licup}},
  \bibinfo {author} {\bibfnamefont {R.}~\bibnamefont {Rens}}, \bibinfo {author}
  {\bibfnamefont {M.}~\bibnamefont {Vahabi}}, \bibinfo {author} {\bibfnamefont
  {K.~A.}\ \bibnamefont {Jansen}}, \bibinfo {author} {\bibfnamefont {G.~H.}\
  \bibnamefont {Koenderink}},\ and\ \bibinfo {author} {\bibfnamefont {F.~C.}\
  \bibnamefont {MacKintosh}},\ }\bibfield  {title} {\bibinfo {title}
  {Strain-driven criticality underlies nonlinear mechanics of fibrous
  networks},\ }\href {https://doi.org/10.1103/PhysRevE.94.042407} {\bibfield
  {journal} {\bibinfo  {journal} {Phys. Rev. E}\ }\textbf {\bibinfo {volume}
  {94}},\ \bibinfo {pages} {042407} (\bibinfo {year}
  {2016}{\natexlab{b}})}\BibitemShut {NoStop}%
\bibitem [{\citenamefont {Vermeulen}\ \emph {et~al.}(2017)\citenamefont
  {Vermeulen}, \citenamefont {Bose}, \citenamefont {Storm},\ and\ \citenamefont
  {Ellenbroek}}]{wouter_pre_2017}%
  \BibitemOpen
  \bibfield  {author} {\bibinfo {author} {\bibfnamefont {M.~F.~J.}\
  \bibnamefont {Vermeulen}}, \bibinfo {author} {\bibfnamefont {A.}~\bibnamefont
  {Bose}}, \bibinfo {author} {\bibfnamefont {C.}~\bibnamefont {Storm}},\ and\
  \bibinfo {author} {\bibfnamefont {W.~G.}\ \bibnamefont {Ellenbroek}},\
  }\bibfield  {title} {\bibinfo {title} {Geometry and the onset of rigidity in
  a disordered network},\ }\href {https://doi.org/10.1103/PhysRevE.96.053003}
  {\bibfield  {journal} {\bibinfo  {journal} {Phys. Rev. E}\ }\textbf {\bibinfo
  {volume} {96}},\ \bibinfo {pages} {053003} (\bibinfo {year}
  {2017})}\BibitemShut {NoStop}%
\bibitem [{\citenamefont {Rens}(2019)}]{rens2019theory}%
  \BibitemOpen
  \bibfield  {author} {\bibinfo {author} {\bibfnamefont {R.}~\bibnamefont
  {Rens}},\ }\href
  {https://dare.uva.nl/search?identifier=6ae60739-28c6-4318-aab9-939a0dacc4bc}
  {\emph {\bibinfo {title} {Theory of rigidity transitions in disordered
  materials}}}\ (\bibinfo  {publisher} {PhD thesis, Univeristy of Amsterdam,
  the Netherlands},\ \bibinfo {year} {2019})\BibitemShut {NoStop}%
\bibitem [{\citenamefont {Wyart}\ \emph {et~al.}(2008)\citenamefont {Wyart},
  \citenamefont {Liang}, \citenamefont {Kabla},\ and\ \citenamefont
  {Mahadevan}}]{maha_prl_2008}%
  \BibitemOpen
  \bibfield  {author} {\bibinfo {author} {\bibfnamefont {M.}~\bibnamefont
  {Wyart}}, \bibinfo {author} {\bibfnamefont {H.}~\bibnamefont {Liang}},
  \bibinfo {author} {\bibfnamefont {A.}~\bibnamefont {Kabla}},\ and\ \bibinfo
  {author} {\bibfnamefont {L.}~\bibnamefont {Mahadevan}},\ }\bibfield  {title}
  {\bibinfo {title} {Elasticity of floppy and stiff random networks},\ }\href
  {https://doi.org/10.1103/PhysRevLett.101.215501} {\bibfield  {journal}
  {\bibinfo  {journal} {Phys. Rev. Lett.}\ }\textbf {\bibinfo {volume} {101}},\
  \bibinfo {pages} {215501} (\bibinfo {year} {2008})}\BibitemShut {NoStop}%
\bibitem [{\citenamefont {Shivers}\ \emph {et~al.}(2019)\citenamefont
  {Shivers}, \citenamefont {Arzash}, \citenamefont {Sharma},\ and\
  \citenamefont {MacKintosh}}]{mackintosh_prl_2019}%
  \BibitemOpen
  \bibfield  {author} {\bibinfo {author} {\bibfnamefont {J.~L.}\ \bibnamefont
  {Shivers}}, \bibinfo {author} {\bibfnamefont {S.}~\bibnamefont {Arzash}},
  \bibinfo {author} {\bibfnamefont {A.}~\bibnamefont {Sharma}},\ and\ \bibinfo
  {author} {\bibfnamefont {F.~C.}\ \bibnamefont {MacKintosh}},\ }\bibfield
  {title} {\bibinfo {title} {Scaling theory for mechanical critical behavior in
  fiber networks},\ }\href {https://doi.org/10.1103/PhysRevLett.122.188003}
  {\bibfield  {journal} {\bibinfo  {journal} {Phys. Rev. Lett.}\ }\textbf
  {\bibinfo {volume} {122}},\ \bibinfo {pages} {188003} (\bibinfo {year}
  {2019})}\BibitemShut {NoStop}%
\bibitem [{\citenamefont {Arzash}\ \emph {et~al.}(2021)\citenamefont {Arzash},
  \citenamefont {Shivers},\ and\ \citenamefont {MacKintosh}}]{arzash2021shear}%
  \BibitemOpen
  \bibfield  {author} {\bibinfo {author} {\bibfnamefont {S.}~\bibnamefont
  {Arzash}}, \bibinfo {author} {\bibfnamefont {J.~L.}\ \bibnamefont
  {Shivers}},\ and\ \bibinfo {author} {\bibfnamefont {F.~C.}\ \bibnamefont
  {MacKintosh}},\ }\bibfield  {title} {\bibinfo {title} {Shear-induced phase
  transition and critical exponents in three-dimensional fiber networks},\
  }\href {https://doi.org/10.1103/PhysRevE.104.L022402} {\bibfield  {journal}
  {\bibinfo  {journal} {Phys. Rev. E}\ }\textbf {\bibinfo {volume} {104}},\
  \bibinfo {pages} {L022402} (\bibinfo {year} {2021})}\BibitemShut {NoStop}%
\bibitem [{\citenamefont {Shivers}\ \emph {et~al.}(2023)\citenamefont
  {Shivers}, \citenamefont {Sharma},\ and\ \citenamefont
  {MacKintosh}}]{fred_arXiv_2022}%
  \BibitemOpen
  \bibfield  {author} {\bibinfo {author} {\bibfnamefont {J.~L.}\ \bibnamefont
  {Shivers}}, \bibinfo {author} {\bibfnamefont {A.}~\bibnamefont {Sharma}},\
  and\ \bibinfo {author} {\bibfnamefont {F.~C.}\ \bibnamefont {MacKintosh}},\
  }\bibfield  {title} {\bibinfo {title} {Strain-controlled critical slowing
  down in the rheology of disordered networks},\ }\href
  {https://doi.org/10.1103/PhysRevLett.131.178201} {\bibfield  {journal}
  {\bibinfo  {journal} {Phys. Rev. Lett.}\ }\textbf {\bibinfo {volume} {131}},\
  \bibinfo {pages} {178201} (\bibinfo {year} {2023})}\BibitemShut {NoStop}%
\bibitem [{\citenamefont {Chen}\ \emph {et~al.}(2023)\citenamefont {Chen},
  \citenamefont {Markovich},\ and\ \citenamefont
  {MacKintosh}}]{chen2023effective}%
  \BibitemOpen
  \bibfield  {author} {\bibinfo {author} {\bibfnamefont {S.}~\bibnamefont
  {Chen}}, \bibinfo {author} {\bibfnamefont {T.}~\bibnamefont {Markovich}},\
  and\ \bibinfo {author} {\bibfnamefont {F.~C.}\ \bibnamefont {MacKintosh}},\
  }\bibfield  {title} {\bibinfo {title} {Effective medium theory for mechanical
  phase transitions of fiber networks},\ }\href
  {https://doi.org/10.1039/D3SM00810J} {\bibfield  {journal} {\bibinfo
  {journal} {Soft Matter}\ }\textbf {\bibinfo {volume} {19}},\ \bibinfo {pages}
  {8124} (\bibinfo {year} {2023})}\BibitemShut {NoStop}%
\bibitem [{\citenamefont {Chen}\ \emph {et~al.}(2024)\citenamefont {Chen},
  \citenamefont {Markovich},\ and\ \citenamefont
  {MacKintosh}}]{freddy_mac_prl_2024}%
  \BibitemOpen
  \bibfield  {author} {\bibinfo {author} {\bibfnamefont {S.}~\bibnamefont
  {Chen}}, \bibinfo {author} {\bibfnamefont {T.}~\bibnamefont {Markovich}},\
  and\ \bibinfo {author} {\bibfnamefont {F.~C.}\ \bibnamefont {MacKintosh}},\
  }\bibfield  {title} {\bibinfo {title} {Field theory for mechanical
  criticality in disordered fiber networks},\ }\href
  {https://doi.org/10.1103/PhysRevLett.133.028201} {\bibfield  {journal}
  {\bibinfo  {journal} {Phys. Rev. Lett.}\ }\textbf {\bibinfo {volume} {133}},\
  \bibinfo {pages} {028201} (\bibinfo {year} {2024})}\BibitemShut {NoStop}%
\bibitem [{\citenamefont {Lerner}\ and\ \citenamefont
  {Bouchbinder}(2023)}]{lerner2023scaling}%
  \BibitemOpen
  \bibfield  {author} {\bibinfo {author} {\bibfnamefont {E.}~\bibnamefont
  {Lerner}}\ and\ \bibinfo {author} {\bibfnamefont {E.}~\bibnamefont
  {Bouchbinder}},\ }\bibfield  {title} {\bibinfo {title} {{Scaling theory of
  critical strain-stiffening in disordered elastic networks}},\ }\href
  {https://doi.org/10.1016/j.eml.2023.102104} {\bibfield  {journal} {\bibinfo
  {journal} {Extreme Mech. Lett.}\ }\textbf {\bibinfo {volume} {65}},\ \bibinfo
  {pages} {102104} (\bibinfo {year} {2023})}\BibitemShut {NoStop}%
\bibitem [{\citenamefont {Arzash}\ \emph {et~al.}(2022)\citenamefont {Arzash},
  \citenamefont {Sharma},\ and\ \citenamefont
  {MacKintosh}}]{freddy_mac_expansion_pre_2022}%
  \BibitemOpen
  \bibfield  {author} {\bibinfo {author} {\bibfnamefont {S.}~\bibnamefont
  {Arzash}}, \bibinfo {author} {\bibfnamefont {A.}~\bibnamefont {Sharma}},\
  and\ \bibinfo {author} {\bibfnamefont {F.~C.}\ \bibnamefont {MacKintosh}},\
  }\bibfield  {title} {\bibinfo {title} {Mechanics of fiber networks under a
  bulk strain},\ }\href {https://doi.org/10.1103/PhysRevE.106.L062403}
  {\bibfield  {journal} {\bibinfo  {journal} {Phys. Rev. E}\ }\textbf {\bibinfo
  {volume} {106}},\ \bibinfo {pages} {L062403} (\bibinfo {year}
  {2022})}\BibitemShut {NoStop}%
\bibitem [{\citenamefont {Lerner}(2024)}]{lerner2024effects}%
  \BibitemOpen
  \bibfield  {author} {\bibinfo {author} {\bibfnamefont {E.}~\bibnamefont
  {Lerner}},\ }\bibfield  {title} {\bibinfo {title} {Effects of coordination
  and stiffness scale separation in disordered elastic networks},\ }\href
  {https://doi.org/10.1103/PhysRevE.109.054904} {\bibfield  {journal} {\bibinfo
   {journal} {Phys. Rev. E}\ }\textbf {\bibinfo {volume} {109}},\ \bibinfo
  {pages} {054904} (\bibinfo {year} {2024})}\BibitemShut {NoStop}%
\bibitem [{\citenamefont {Lerner}\ \emph {et~al.}(2014)\citenamefont {Lerner},
  \citenamefont {DeGiuli}, \citenamefont {During},\ and\ \citenamefont
  {Wyart}}]{breakdown}%
  \BibitemOpen
  \bibfield  {author} {\bibinfo {author} {\bibfnamefont {E.}~\bibnamefont
  {Lerner}}, \bibinfo {author} {\bibfnamefont {E.}~\bibnamefont {DeGiuli}},
  \bibinfo {author} {\bibfnamefont {G.}~\bibnamefont {During}},\ and\ \bibinfo
  {author} {\bibfnamefont {M.}~\bibnamefont {Wyart}},\ }\bibfield  {title}
  {\bibinfo {title} {Breakdown of continuum elasticity in amorphous solids},\
  }\href {https://doi.org/10.1039/C4SM00311J} {\bibfield  {journal} {\bibinfo
  {journal} {Soft Matter}\ }\textbf {\bibinfo {volume} {10}},\ \bibinfo {pages}
  {5085} (\bibinfo {year} {2014})}\BibitemShut {NoStop}%
\bibitem [{\citenamefont {Kapteijns}\ \emph {et~al.}(2021)\citenamefont
  {Kapteijns}, \citenamefont {Bouchbinder},\ and\ \citenamefont
  {Lerner}}]{quantifier_PRE_2021}%
  \BibitemOpen
  \bibfield  {author} {\bibinfo {author} {\bibfnamefont {G.}~\bibnamefont
  {Kapteijns}}, \bibinfo {author} {\bibfnamefont {E.}~\bibnamefont
  {Bouchbinder}},\ and\ \bibinfo {author} {\bibfnamefont {E.}~\bibnamefont
  {Lerner}},\ }\bibfield  {title} {\bibinfo {title} {Unified quantifier of
  mechanical disorder in solids},\ }\href
  {https://doi.org/10.1103/PhysRevE.104.035001} {\bibfield  {journal} {\bibinfo
   {journal} {Phys. Rev. E}\ }\textbf {\bibinfo {volume} {104}},\ \bibinfo
  {pages} {035001} (\bibinfo {year} {2021})}\BibitemShut {NoStop}%
\bibitem [{\citenamefont {Lutsko}(1989)}]{lutsko1989generalized}%
  \BibitemOpen
  \bibfield  {author} {\bibinfo {author} {\bibfnamefont {J.}~\bibnamefont
  {Lutsko}},\ }\bibfield  {title} {\bibinfo {title} {Generalized expressions
  for the calculation of elastic constants by computer simulation},\ }\href
  {https://doi.org/10.1063/1.342716} {\bibfield  {journal} {\bibinfo  {journal}
  {Journal of applied physics}\ }\textbf {\bibinfo {volume} {65}},\ \bibinfo
  {pages} {2991} (\bibinfo {year} {1989})}\BibitemShut {NoStop}%
\bibitem [{\citenamefont {Bitzek}\ \emph {et~al.}(2006)\citenamefont {Bitzek},
  \citenamefont {Koskinen}, \citenamefont {G\"ahler}, \citenamefont {Moseler},\
  and\ \citenamefont {Gumbsch}}]{bitzek2006structural}%
  \BibitemOpen
  \bibfield  {author} {\bibinfo {author} {\bibfnamefont {E.}~\bibnamefont
  {Bitzek}}, \bibinfo {author} {\bibfnamefont {P.}~\bibnamefont {Koskinen}},
  \bibinfo {author} {\bibfnamefont {F.}~\bibnamefont {G\"ahler}}, \bibinfo
  {author} {\bibfnamefont {M.}~\bibnamefont {Moseler}},\ and\ \bibinfo {author}
  {\bibfnamefont {P.}~\bibnamefont {Gumbsch}},\ }\bibfield  {title} {\bibinfo
  {title} {Structural relaxation made simple},\ }\href
  {https://doi.org/10.1103/PhysRevLett.97.170201} {\bibfield  {journal}
  {\bibinfo  {journal} {Phys. Rev. Lett.}\ }\textbf {\bibinfo {volume} {97}},\
  \bibinfo {pages} {170201} (\bibinfo {year} {2006})}\BibitemShut {NoStop}%
\bibitem [{\citenamefont {Calladine}(1978)}]{Calladine_1978}%
  \BibitemOpen
  \bibfield  {author} {\bibinfo {author} {\bibfnamefont {C.}~\bibnamefont
  {Calladine}},\ }\bibfield  {title} {\bibinfo {title} {Buckminster fuller's
  ``tensegrity" structures and clerk maxwell's rules for the construction of
  stiff frames},\ }\href
  {https://doi.org/https://doi.org/10.1016/0020-7683(78)90052-5} {\bibfield
  {journal} {\bibinfo  {journal} {Int. J. Solids Struct.}\ }\textbf {\bibinfo
  {volume} {14}},\ \bibinfo {pages} {161} (\bibinfo {year} {1978})}\BibitemShut
  {NoStop}%
\bibitem [{\citenamefont {D\"uring}\ \emph
  {et~al.}(2014{\natexlab{b}})\citenamefont {D\"uring}, \citenamefont
  {Lerner},\ and\ \citenamefont {Wyart}}]{during2014length}%
  \BibitemOpen
  \bibfield  {author} {\bibinfo {author} {\bibfnamefont {G.}~\bibnamefont
  {D\"uring}}, \bibinfo {author} {\bibfnamefont {E.}~\bibnamefont {Lerner}},\
  and\ \bibinfo {author} {\bibfnamefont {M.}~\bibnamefont {Wyart}},\ }\bibfield
   {title} {\bibinfo {title} {Length scales and self-organization in dense
  suspension flows},\ }\href {https://doi.org/10.1103/PhysRevE.89.022305}
  {\bibfield  {journal} {\bibinfo  {journal} {Phys. Rev. E}\ }\textbf {\bibinfo
  {volume} {89}},\ \bibinfo {pages} {022305} (\bibinfo {year}
  {2014}{\natexlab{b}})}\BibitemShut {NoStop}%
\bibitem [{\citenamefont {Rens}\ \emph
  {et~al.}(2018{\natexlab{b}})\citenamefont {Rens}, \citenamefont {Villarroel},
  \citenamefont {D{\"u}ring},\ and\ \citenamefont
  {Lerner}}]{rens2018micromechanical}%
  \BibitemOpen
  \bibfield  {author} {\bibinfo {author} {\bibfnamefont {R.}~\bibnamefont
  {Rens}}, \bibinfo {author} {\bibfnamefont {C.}~\bibnamefont {Villarroel}},
  \bibinfo {author} {\bibfnamefont {G.}~\bibnamefont {D{\"u}ring}},\ and\
  \bibinfo {author} {\bibfnamefont {E.}~\bibnamefont {Lerner}},\ }\bibfield
  {title} {\bibinfo {title} {Micromechanical theory of strain stiffening of
  biopolymer networks},\ }\href {https://doi.org/10.1103/PhysRevE.98.062411}
  {\bibfield  {journal} {\bibinfo  {journal} {Phys. Rev. E}\ }\textbf {\bibinfo
  {volume} {98}},\ \bibinfo {pages} {062411} (\bibinfo {year}
  {2018}{\natexlab{b}})}\BibitemShut {NoStop}%
\end{thebibliography}

%

\end{document}